\documentclass[useAMS,usenatbib]{mn2e}

\usepackage{graphicx}
\usepackage{lscape}
\usepackage{rotating}
\usepackage{booktabs}
\usepackage{longtable}
\usepackage{multicol} 

\title[Studying the asymmetry of the GC population of NGC~4261]{Studying the asymmetry of the GC population of NGC~4261}

\author[P.~Bonfini et al.]
{
 \parbox{\textwidth}{
  P.~Bonfini,$^{1,2}$\thanks{E-mail: \texttt{paolo@physics.uoc.gr}}
  A. Zezas,$^{1,2,3}$
  M. Birkinshaw,$^{4}$
  D.~M. Worrall,$^{4}$
  G. Fabbiano,$^{3}$
  E. O'Sullivan,$^{5,3}$
  G. Trinchieri$^{6}$
  and
  A. Wolter$^{6}$
 }
 \vspace{0.4cm}\\
 \parbox{\textwidth}{
  $^{1}$University of Crete, Physics Department \& Institute of Theoretical \& Computational Physics, 71003 Heraklion, Crete, Greece\\
  $^{2}$Foundation for Research and Technology-Hellas, 71110 Heraklion, Crete, Greece\\
  $^{3}$Center for Astrophysics, 60 Garden Street, Cambridge, MA 02138\\
  $^{4}$H.H. Wills Physics Laboratory, University of Bristol, Tyndall Avenue, Bristol BS8 1TL,  U.K.\\
  $^{5}$School of Physics and Astronomy, University of Birmingham, Birmingham, B15 2TT, U.K. \\
  $^{6}$INAF, Osservatorio Astronomico di Brera, Via Brera 28 IT 20121, Milano, Italy\\
 }
}

\begin{document}

\date{Accepted 2012 January 7. Received 2011 December 29; in original form 2011 September 13}

\pagerange{000--000} \pubyear{0000}

\maketitle

\label{firstpage}

\begin{abstract}

We present an analysis of the Globular Cluster (GC) population of the elliptical
galaxy NGC~4261 based on $HST$ WFPC2 data in the $B$, $V$ and $I$ bands.
We study the spatial distribution of the GCs in order to probe the anisotropy
in the azimuthal distribution of the discrete X-ray sources in the galaxy revealed by
\emph{Chandra} images \citep{zez03}.
The luminosity function of our GC sample (complete at the 90\% level for
$m_{_{V}}$ = 23.8~mag) peaks at $m_{_{V}}$ = 25.1$_{-0.6}^{+1.0}$~mag, which
corresponds to a distance consistent with previous measurements.
The colour distribution can be interpreted as being the
superposition of a blue and red GC component with average colours
$V-I$ = 1.01$_{-0.06}^{+0.06}$~mag and 1.27$_{-0.08}^{+0.06}$~mag, respectively.
This is consistent with a bimodal colour distribution typical of elliptical
galaxies.
The red GC's radial profile is steeper than that of the galaxy surface brightness,
while the profile of the blue subpopulation looks more consistent with it.
The most striking finding is the significant asymmetry in the azimuthal distribution
of the GC population about a NE-SW direction.
The lack of any obvious feature in the morphology of the galaxy suggests that the
asymmetry could be the result of an interaction or a merger.

\end{abstract}

\begin{keywords}
globular clusters: general --- galaxies: individual (NGC~4261, 3C~270) --- galaxies: interactions
\end{keywords}

\section{Introduction}
\label{Introduction}

Elliptical galaxies are spheroidal systems which are expected to present a highly
uniform spatial distribution of starlight and Globular Cluster (GC) sources.
Nevertheless, optical images show that several elliptical galaxies exhibit structures
such as shells, ripples, arcs, tidal tails and other asymmetric features \citep[e.g.][]{sch92}.
One of the most popular scenarios for elliptical galaxy formation suggests
that these features are the result of major mergers.
Numerical simulations \citep[e.g.][]{too72,bar88} show indeed that such mergers are
characterized by peculiar remnant structures of this type.
Minor merging events and interactions can subsequently disturb further the galaxy
morphology.

Merger events and galaxy encounters affect the spatial distribution of the stellar
population as well as that of the GC population.
\cite{sch92} discovered significant structure in the starlight of several elliptical
galaxies, and disturbed orbits are found in NGC1404 and NGC1399, probably due to
a dynamical interaction between the two galaxies (\citealt{nap02,bek03}).
The amplitude of the deviation from the isotropy in either
the stellar or GC populations in elliptical systems obviously depends on the time lapsed
since the merging/interaction event, as well as on the nature of the event itself,
and so may be used to explore the elliptical galaxy's formation history.

\subsection{The Case of NGC4261}
\label{The Case of NGC4261}
NGC~4261 is a nearby (29.4$\pm$2.6~Mpc; \citealt{jen03}) early-type galaxy in the Virgo
W cloud \citep{gar93,nol93}.
It is a well-studied object thanks to many interesting
features: it has an active nucleus hosting a supermassive black hole
(4.9$\pm$1.0)$\times$10$^{8}$M$_{\sun}$ \citep{fer96}, it hosts the radio source 3C~270,
whose prominent radio jets \citep{bir85} are associated with X-ray emission close to the
nucleus \citep{wor10}. It shows a 20\arcsec dust lane along the north-south axis \citep{mar00}
and has an impressive nuclear dust disk, discovered with \emph{HST} WFPC2 \citep{jaf96}.

Apart from the ``boxy`` isophotes \citep[e.g.][]{nie89} and weak evidence for a
tidal arm in the NW direction \citep{tal09},
this galaxy does not show any definite story of recent interaction such as shells,
ripples, rings, resulting in a low fine structure parameter \citep[$\Sigma$=1.0;][]{sch92}.
Therefore, despite being the most massive object in a poor group
of galaxies \citep{dav95}, we can claim that NGC~4261 does not show evidence of
recent gravitational interaction.

Given the undisturbed starlight distribution of NGC~4261, the discrete
X-ray sources (i.e. X-ray binaries) would be expected to be distributed uniformly.
However, \emph{Chandra} data pointed to an interesting anisotropy in their azimuthal
distribution, and more precisely to an excess between P.A.=140$^{\circ}$ and
P.A.=190$^{\circ}$ \citep{zez03}.
\cite{gio05} associated 50\% of these sources with GCs identified in
archival optical images and suggested that the \emph{Chandra} sources are
accreting Low Mass X-ray Binaries (LMXBs).
This result suggests a non-uniform distribution of the GC population.

In order to investigate this hypothesis, we requested deep \emph{HST} WFPC2 data
(proposal ID~11339; PI: Zezas, A - Table \ref{log}) to characterize the GC
population of NGC~4261. In this paper we present a study
of the radial and azimuthal distributions of the blue (metal poor) and red (metal rich)
GC subpopulations. We assess the extent of the asymmetry and discuss its origin
in the context of a galaxy interaction.
In a following paper (Bonfini et al., in preparation), we will extend the
analysis to the LMXB population using the new deep \emph{Chandra} pointing.
In particular, by probing whether the LMXBs and
GCs indeed have similar spatial distribution, we will add to the debate on
whether GCs are the sole birthplaces of LMXBs \citep[e.g.][]{whi02}, or field LMXBs
form independently in situ \citep[e.g.][and references therein]{kun07}, or, as
suggested by the recent work of \cite{kim09}, both formation processes
contribute to the field LMXB population of elliptical galaxies.

The outline of the paper is as follows.
In $\S$\ref{Optical Observations and Data Analysis} we describe the \emph{HST}
data reduction and source detection.
In $\S$\ref{Definition of the Globular Cluster Population}
we report the criteria used to define GC candidates and we
describe the procedure used to correct the Luminosity Function (LF) for incompleteness.
In $\S$\ref{Properties of the Globular Clusters} we study the
Globular Cluster Luminosity Function (GCLF) and colour distribution.
In $\S$\ref{Spatial Distribution of Sources}
we study the azimuthal and radial distributions of the GCs.
In $\S$\ref{Discussion}
we discuss our results in the context of the recent interaction
history of NGC~4261.

\begin{figure}
 \includegraphics[width=0.48\textwidth,angle=0]{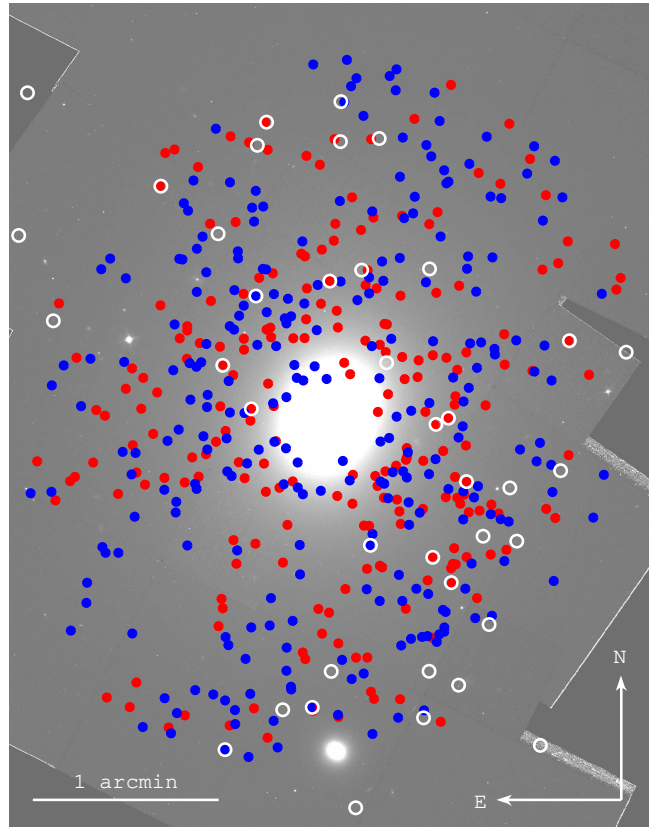}
 \caption{
  $I$ band mosaic. 
  Blue and red markers represent the position of
  blue and red secure GCs down to the 75\% completeness limit
  (sample and choice of completeness limit are defined in Section
  $\S$\ref{Definition of the Globular Cluster Population} and
  $\S$\ref{Spatial Distribution of Sources}, respectively). These
  are the sources on which we performed the study of the spatial distribution.
  The white open circles represent the locations of X-ray sources in the
  source list of \citealt{gio05}.
  \label{mosaic}
 }
\end{figure}

\section{Optical Observations and Data Analysis}
\label{Optical Observations and Data Analysis}

\subsection{Data Reduction}
\label{Data Reduction}
\emph{HST} WFPC2 data for NGC~4261 were obtained with the F450W, F606W and F814W
filters (roughly corresponding to the B, V and I bands, respectively;
see WFPC2 Instrument Handbook, Version 10.0,
Table 3.1\footnote{
 http://www.stsci.edu/hst/wfpc2/documents/IHB$\_$17.html}
).
The filters were chosen to provide the highest sensitivity while allowing
us to distinguish the blue and red GC subpopulations.

In order to reject cosmic rays the observations were performed
following a two point dither pattern with a 0.3 pixel offset.
A total of 12 exposures of 400~s each were acquired in each filter in a
3$\times$2 grid.
The pointings were planned in order to cover the $D_{25}$
\citep[major diameter; 4\arcmin.07][]{dev91} area of the galaxy.
The details of the observations are reported in 
Table \ref{log}.

\begin{table}
 \centering
  \caption{L{\tiny OG} O{\tiny F} T{\tiny HE} O{\tiny BSERVATIONS}\label{log}}
  \begin{tabular}{@{}ccc@{}cc@{}}
   \hline
   Instrument & Filter &  Number   & Date &  Exposure \\
              &        &  of       &      & per Field\\
              &        & Fields    &      & (sec)\\
   \hline
   \emph{HST} WFPC2 & F450W ($B$) & 6 & 2008 Jan 12 & (2$\times$)400\\
                    &             &   & 2008 Jan 13 &               \\
   \emph{HST} WFPC2 & F606W ($V$) & 6 & 2007 Dec 26 & (2$\times$)400\\
   \emph{HST} WFPC2 & F814W ($I$) & 6 & 2008 Jan 28 & (2$\times$)400\\
                    &             &   & 2008 Mar 10 &               \\
   \hline
  \end{tabular}
\end{table}

Mosaics of the HST images were created with the following procedure.
The task \emph{MultiDrizzle} (within the {\footnotesize PyRAF} package,
DITHER version 2.3)
had been first used to produce the distortion corrected frames for each exposure.
Then, the relative offsets between the fields were determined by comparing the
coordinates of stars present in overlapping regions of these frames.
The pointlike sources within each frame were detected using
\emph{daofind}, while the {\footnotesize PyRAF} tasks \emph{xyxymatch} and
\emph{geomap} were used to calculate the offsets.
Due to the small number of stars in the field of view, we could
identify only 2-5 sources in common between each pair of fields, therefore in
some cases we had to include some GCs as reference objects.
The offset table created in this way was fed to \emph{MultiDrizzle} in order
to create the final version of the mosaic.
The mosaics for the different filters were registered with respect to the
$I$ band image.
The final image was binned to the Wide Field Camera CCD pixel scale
(WFC; nominally, 0.0996\arcsec/pixel).

Subsequently, we corrected for systematic astrometry errors by
cross-correlating the HST coordinates of the 50 brightest objects (counts~$>$~4000)
in the $I$ band mosaic (see $\S$\ref{Source Identification and Photometry}
for the source detection technique) against the coordinates of the 50 brightest
\emph{Sloan Digital Sky Survey - Data Release 5} \citep[SDSS DR5;][]{sdss} sources
in the field. The SDSS catalogue provided a more reliable calibration since it is richer
in astrometric reference stars than the HST field.
We performed the astrometric correction using the {\footnotesize WCSTOOLS} \emph{imwcs},
which produced a significant number of matches (11/50), allowing us to measure
a negligible \emph{WCS} offset of
\mbox{($\Delta$RA,$\Delta$Dec) = (-0.02$\pm$0.10\arcsec,0.01$\pm$0.13\arcsec)}.
The same \emph{WCS} solution was then applied to the $B$ and $V$ band mosaics.

Figure \ref{mosaic} shows the final drizzled image
for filter $I$, along with the location of the blue and red GCs of our secure sample
down to the 75\% completeness limit (the sample and choice of completeness limit are defined
in $\S$\ref{Definition of the Globular Cluster Population} and
$\S$\ref{Spatial Distribution of Sources}, respectively).
In the figure we also show the locations of the X-ray sources
in the list of \cite{gio05}.

\subsection{Source Identification and Photometry}
\label{Source Identification and Photometry}
We used SE{\footnotesize XTRACTOR} \citep{ber96} to detect discrete sources in the mosaics.
This package performs source detection and provides net counts for each
source after estimating and subtracting the local background.
Following \cite{for04}, who used SE{\footnotesize XTRACTOR}
for the detection of GCs in {\footnotesize WFPC2} data of nearby galaxies,
we chose the following software parameters: minimum detection area
of 4 pixels; detection threshold of 1.5~$\sigma$;
threshold for photometry analysis of 1.5~$\sigma$; background grid of
16$\times$16 pixels; filtering Gaussian of 2.5 pixels Full Width at Half Maximum (FWHM;
roughly equal to the value of the FWHM of the point-like sources in the mosaics).

A typical GC has a half light radius \emph{r$_{h}$}$\simeq$3~pc
\citep[e.g.][]{ash98}.
At the distance of NGC~4261 this value translates into a size of 0.02\arcsec or 0.2 pixels
at the scale of our mosaics, meaning that the GCs in our images (including
the wings in their light profile) are only marginally resolved,
which does not allow accurate comparison to the Point Spread Function
(PSF; PSF FWHM $\sim$ 2 pixels in our mosaics) to determine their extent.
Considering the low luminosity of the sources, we decided to perform fixed aperture
photometry in order to minimize the photometric error.
This decision is justified by the results of the incompleteness simulation
(see $\S$\ref{Evaluating Incompleteness}).
We adopted an aperture diameter of 5 pixels (or 0.5$\arcsec$), encompassing $\sim$95\%
of the total flux of a point source \citep{hol_a}.

The conversion from instrumental magnitudes to the $UBVRI$ system was performed
using the WFPC2 calibrations of \cite{hol_a}. When available, transformation
parameters based on observational data were preferred over synthetic ones.
The $B$-$V$ and $V$-$I$ colours were obtained from an analytic solution of equation (8)
of \cite{hol_a} for the single $B$, $V$ and $I$ bands:

\begin{tiny}
\begin{equation}
 B-V =  - { { (T_{1,FS}^{B}-T_{1,FS}^{V}-1) + \sqrt{\Delta_{BV} } }\over{ 2(T_{2,FS}^{B}-T_{2,FS}^{V}) } }
\end{equation}
\begin{equation}
 V-I =  - { { (T_{1,FS}^{V}-T_{1,FS}^{I}-1) + \sqrt{\Delta_{VI} } }\over{ 2(T_{2,FS}^{V}-T_{2,FS}^{I}) } }
\end{equation}
\end{tiny}
with:
\begin{tiny}
\begin{equation}
 \Delta_{BV} =  (T_{1,FS}^{B}-T_{1,FS}^{V}-1)^{2} - 4(T_{2,FS}^{B}-T_{2,FS}^{V})(-2.5log{ f_{B} \over f_{V} } + (Z_{FS}^{B}-Z_{FS}^{V}))
\end{equation}
\begin{equation}
 \Delta_{VI} =  (T_{1,FS}^{V}-T_{1,FS}^{I}-1)^{2} - 4(T_{2,FS}^{V}-T_{2,FS}^{I})(-2.5log{ f_{V} \over f_{I} } + (Z_{FS}^{V}-Z_{FS}^{I}))
\end{equation}
\end{tiny}

\noindent where $T_{1,FS}$ and $T_{2,FS}$ are the flight-to-$UBVRI$ transformation coefficients,
$Z_{FS}$ the zeropoints from \cite{hol_a} and $f_{B}$, $f_{V}$ and $f_{I}$ correspond to the aperture
fluxes in the $B$, $V$ and $I$ band respectively.

Aperture corrections (evaluated with the method described in
$\S$\ref{Evaluating Incompleteness}) were applied to account for aperture losses.

\section{Definition of the Globular Cluster Population}
\label{Definition of the Globular Cluster Population}

\subsection{Globular Cluster Selection}
\label{Globular Cluster Selection}
We created our GCs list by cross-correlating the SE{\footnotesize XTRACTOR}
output catalogues for the $V$ and $I$ band.
The $B$ band catalogue was excluded from the cross-correlation due to the limited
number of sources detected in this band (because of the relatively red colour of GCs).
The result of the match was subsequently screened according to the following criteria:

 \emph{Selection on position.} We restricted the sample to sources within the
 $D_{25}$ ellipse of the galaxy (4\arcmin.07 major diameter; 0.8 axis ratio $b/a$).
 The SE{\footnotesize XTRACTOR} source detection efficiency dropped drastically at the
 galaxy center due to the high background, with virtually no detections within a
 galactocentric radius of $R_{GC}$ = 25$\arcsec$.

 \emph{Selection on axial ratio.} GCs are usually close to perfect
 spherical systems, therefore we could confidently exclude objects
 with significant axial ratios. As in \cite{jor04} we adopted a mean
 axial ratio $\langle\in\rangle$ (i.e., the average of the axial ratios in
 different filters) and we set conservative limits 0.5 $<\langle\in\rangle<$ 2.0.

 \emph{Selection on $\langle$S/N$\rangle$.} We limited the signal to noise
 ratio of each source to be higher than 20 for both the $V$ and $I$
 band\footnote{
  The SE{\footnotesize XTRACTOR} thresholds discussed above
  refer to the $\langle$S/N$\rangle$ for any individual pixel of a source,
  rather than to the $\langle$S/N$\rangle$ for the total flux within the aperture.
 }.

 \emph{Selection on FWHM.} As mentioned in Section
 \ref{Source Identification and Photometry}, GCs at the distance of NGC~4261 are
 expected to be only marginally resolved. Therefore, the minimum  FWHM is dictated
 by the PSF FWHM of the HST WFPC2 camera, which is around 2-2.5 pixels
 \citep[see][Figure 5]{hol_b}.
 Quantifying the upper limit, instead, is a more difficult task since it
 should take into account the convolution of the radial profile of an extended GC
 with the instrument PSF. Based on extensive simulations 
 (see $\S$\ref{Evaluating Incompleteness}), we set the upper limit to a
 few times the PSF FWHM in order to account for the largest GCs.
 Larger objects are expected to be
 background galaxies and therefore must be rejected.
 Based on these considerations, we restricted the FWHM within the limits:
 1.5$<$FWHM$<$4.0~pixels, for every band.

\emph{Selection on colour.} The colour was constrained
 to be in the range observed in elliptical galaxy GCs:
 0.6$<$$V$-$I$$<$1.6 \citep[e.g.][]{ash98,kun01}.

The forementioned criteria resulted in a sample of 718 ``secure GCs''
(down to a minimum $V$ magnitude of $\sim$25.4~mag).
A second, less strict sample, was defined from the SE{\footnotesize XTRACTOR}
catalogue for the $V$ band applying the same selection except for the colour criterion.
In this way, we found a total of 1067 ``GC candidates''
(down to a minimum $V$ magnitude of $\sim$25.8~mag).

False detections associated with background galaxies, although not excluded, are
expected to be minimal, due to the restrictions applied on the axis ratio and FWHM and,
in the case of the secure sample, to the colour limits (late type galaxies
are much bluer than the typical GC).
As a further test, we investigated the effectiveness of the FWHM selection
in rejecting background galaxies as opposed to an equivalent criterium often used
in similar HST studies of extragalactic GCs, which is the difference in magnitude
within two different apertures.
To do so, we performed a comparison of the flux ratio between a 1.5 pixels and a 2.5
pixels aperture (roughly the size of the PSF FWHM) against the FWHM of the candidate
GCs sample. This comparison showed a tight correlation between the two quantities (although
affected by significant scatter).
Therefore, applying either of the two criteria would lead to the same results.
We also verified that source confusion (i.e. blending of sources) is negligible,
if present at all.

Table \ref{catalogue} lists the photometric parameters of the objects in the secure
GCs sample. When available, the $B$ magnitude and the $B$-$V$ colour have been included.

\subsection{Evaluating Incompleteness}
\label{Evaluating Incompleteness}

\begin{figure*}
 \makebox[\linewidth]{
  \includegraphics[width=0.48\textwidth,angle=0]{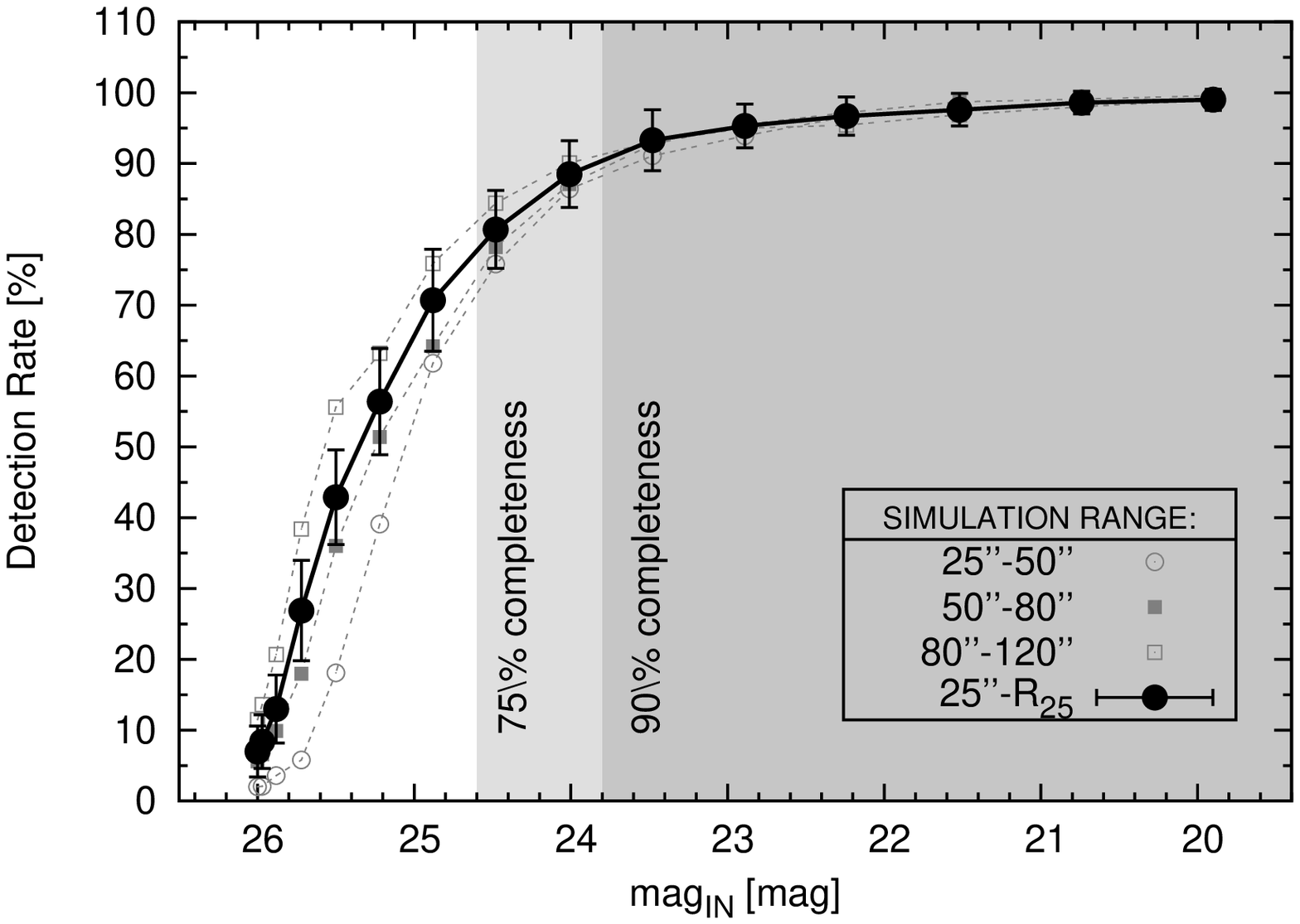}
  \includegraphics[width=0.48\textwidth,angle=0]{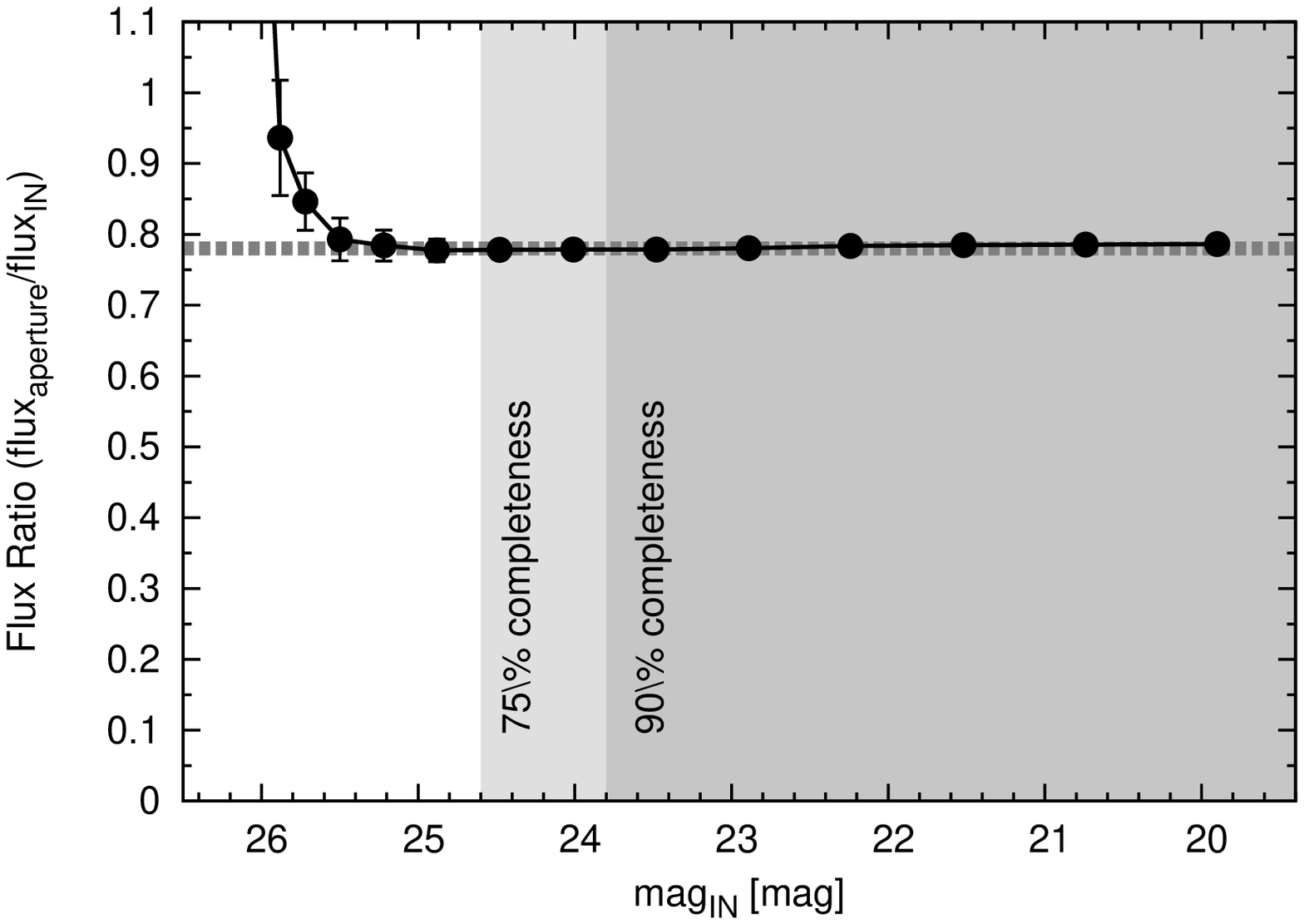}
 }
 \makebox[\linewidth]{
  \includegraphics[width=0.48\textwidth,angle=0]{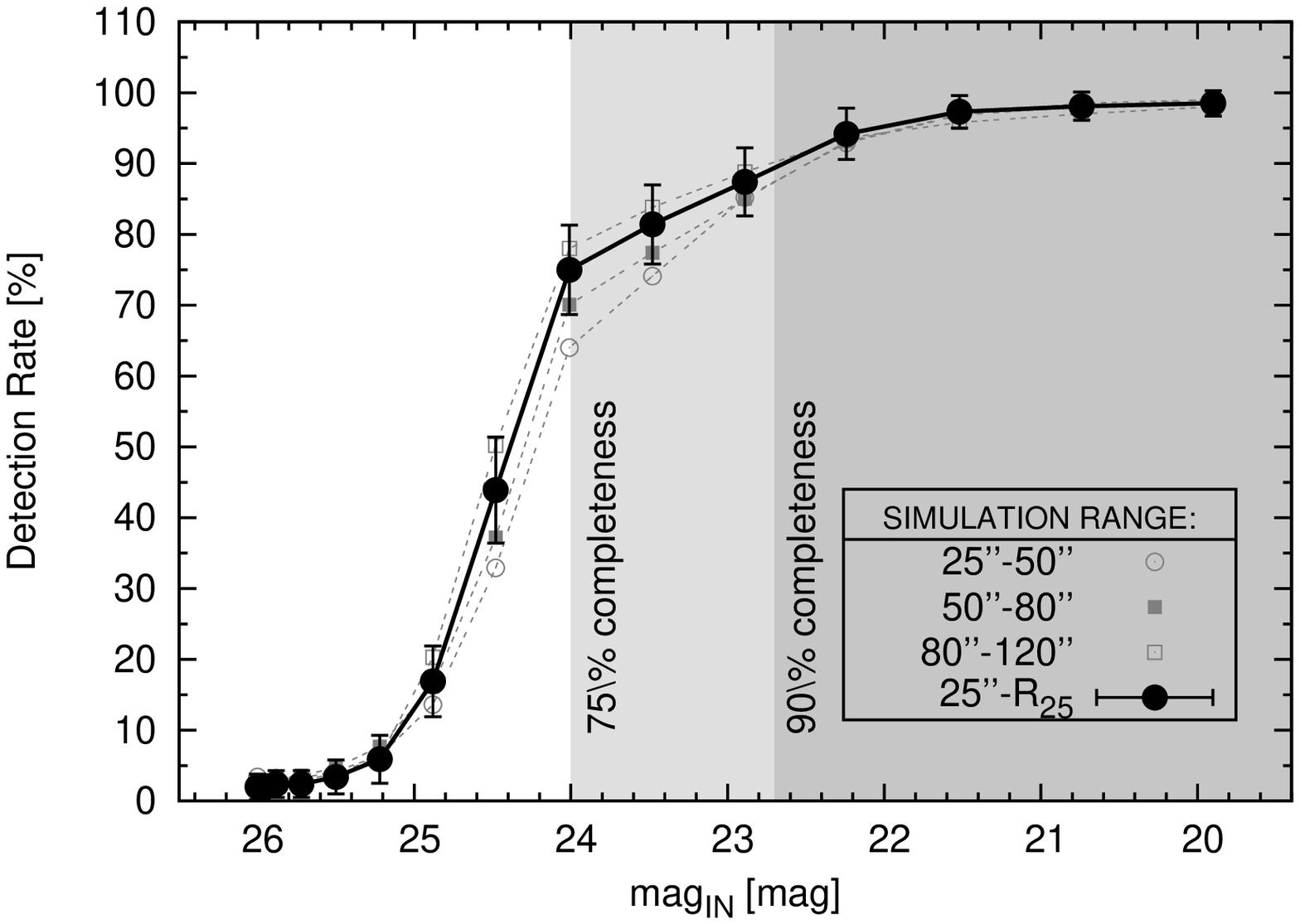}
  \includegraphics[width=0.48\textwidth,angle=0]{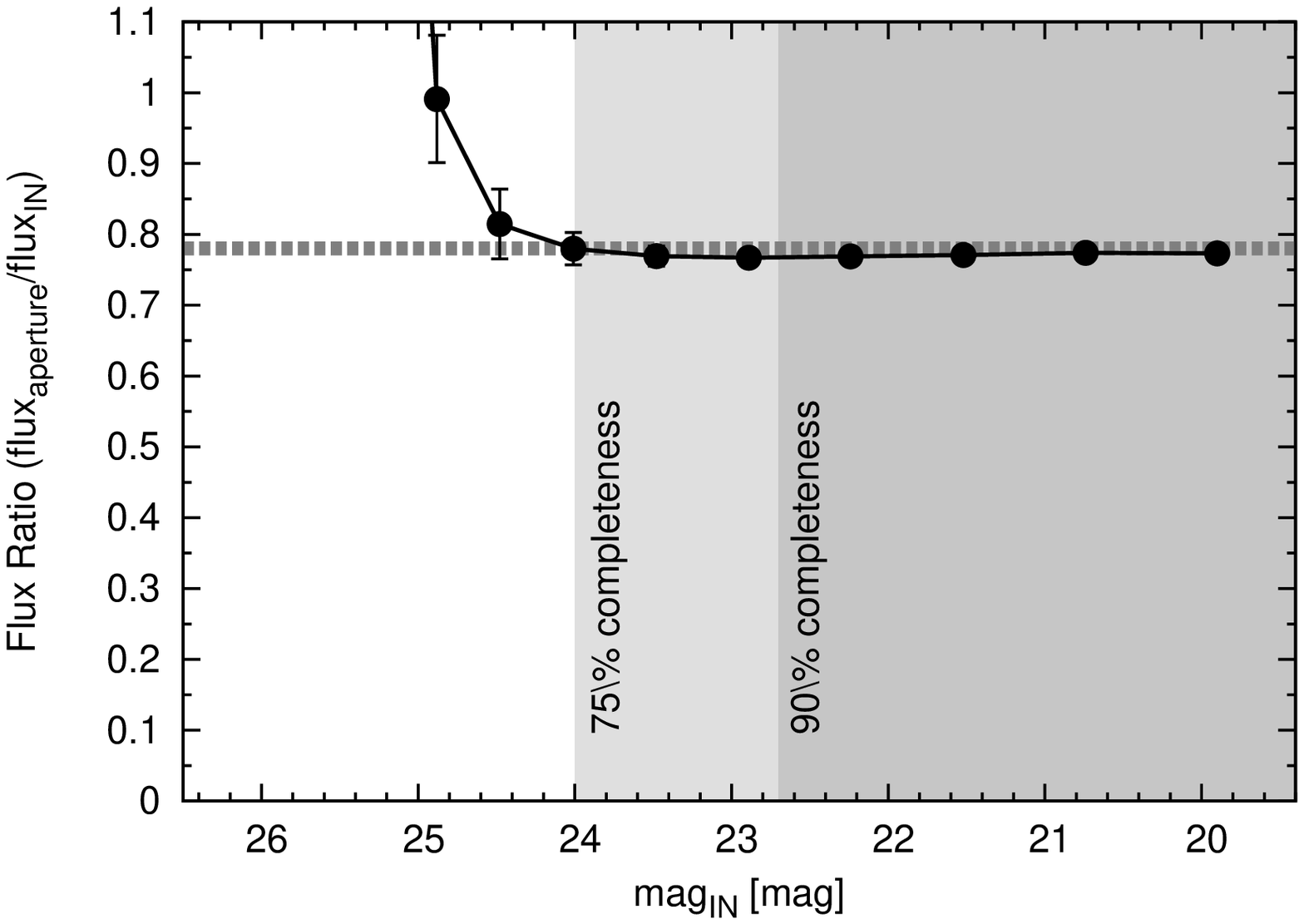}
 }
 \caption{
 Results of the artificial source simulation for the $V$ (top) and $I$ (bottom) bands.
 The results were screened using the same criteria used to define the ``GC candidates''
 sample ($\S$\ref{Globular Cluster Selection}).
 \emph{Left.}- Detection rate (i.e. percentage of detected sources over simulated sources)
 as a function of the input magnitude. The shadowed areas delimit the 90\% and 75\%
 completeness regions. The bold line represents the detection rate for the simulation run
 between the detection limit ($R_{GC} = 25~\arcsec$) and the galaxy $R_{25}$. The light
 gray lines represent the results from simulations run within limited galactocentric ranges.
 The dependence of the detection rate curves to the radial distance from the galaxy center
 has been properly taken into account when correcting the $V$ magnitude histogram for the
 effects of incompleteness (see $\S$\ref{$V$ band Luminosity Function}).
 \emph{Right.}- Ratio of the aperture flux measured by SE{\tiny XTRACTOR}
 over the total input flux of the simulated source, as a function of the input magnitude.
 The median value for the data points over 75\% completeness, represented by the dashed line,
 has been used to compute the aperture corrections applied
 to the fluxes of the real GCs (see $\S$\ref{Source Identification and Photometry}).
 \label{incompleteness}
}
\end{figure*}

In order to evaluate the incompleteness of the GC samples
(i.e., the fraction of GCs not detected due to faintness or issues related to the detection
process),
we set up an artificial source test to calibrate the SE{\footnotesize XTRACTOR} results.
Simulated GCs were added to the NGC~4261 \emph{HST} mosaics and their
characteristics were measured with SE{\footnotesize XTRACTOR} using the same
setup as for the real data. The simulation was repeated several times in order
to improve the statistical results.
The details of the artificial source test are described in
\mbox{Appendix \ref{INCOMPLETENESS SIMULATION}}.

\smallskip

Figure \ref{incompleteness} reports the simulation results
for the $V$ (top) and $I$ (bottom) bands. The left panels of the figure show
the detection rate (i.e. percentage of detected sources over simulated sources) as
a function of the input magnitude.
Input sources are detected with an efficiency of $\sim$90\% (or higher)
down to 23.8~mag and 22.7~mag for the $V$ and $I$ band respectively.
We define these magnitudes as the detection thresholds.
In order to assess the dependence of the completeness curves on the background light
we repeated the simulation within 3 annulii located at different galactocentric radii
(between the simulation limits $R_{GC}$ = 25$\arcsec$ and $R_{GC}$ = $R_{25}$).
We found that the detection efficiency indeed shows a dependence on the brightness
of the background, although the differences are significant only well below the 80\%
completeness level. The completeness correction applied to the $V$ band histogram
in $\S$\ref{$V$ band Luminosity Function} took this effect into account.
Since the detection efficiency drops drastically for $R_{GC} < 25~\arcsec$, we
adopted this as the minimum galactocentric radius for the study of the spatial
distribution of the sources ($\S$\ref{Spatial Distribution of Sources}).
The right panels of Figure \ref{incompleteness} show the ratio of the 0.5$\arcsec$ diameter
aperture flux measured with SE{\footnotesize XTRACTOR} over the total input flux of the simulated
sources, as a function of the input magnitude.
At the faintest magnitudes, the flux ratio deviates from the constant value since
the Poissonian variation of the background counts becomes an important contribution
to the signal within the aperture. For this reason, we estimated the aperture correction
only down to the 75\% completeness limit.
We find that our fixed 0.5$\arcsec$ aperture encompasses 79\%, 78\% and 77\% of the flux
for the $B$, $V$ and $I$ band respectively regardless of the size and shape parameters
of the sources.
These ratios are used as the aperture corrections for the magnitudes of the observed GCs
(see $\S$\ref{Source Identification and Photometry}).

\begin{figure}
 \includegraphics[width=0.48\textwidth,angle=0]{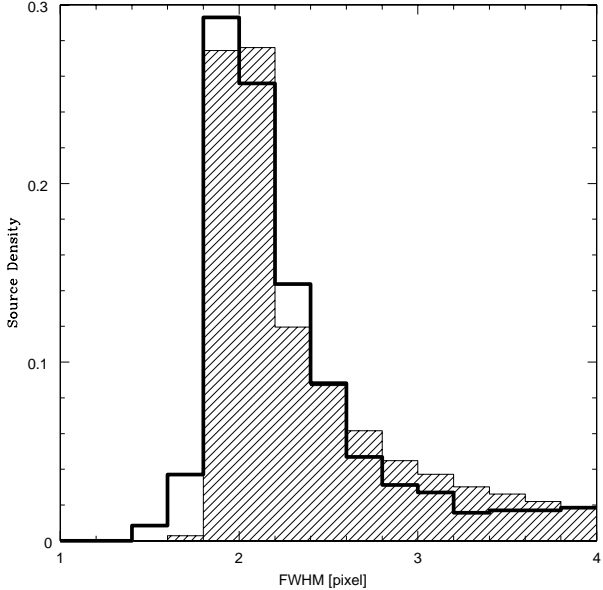}
 \caption{
  Comparison of the FWHMs (in units of WFPC2 pixels) of the simulated sources
  (solid line) versus the FWHMs of the measured sources in the data (shaded histogram)
  for the $V$ band. The histograms are normalized to the total number of sources with
  FWHM between 0 and 4 pixels. This result is valid for the ``GC candidates'' sample
  (no colour selection was applied in this simulation).
  \label{FWHM}
 }
\end{figure}

Figure \ref{FWHM} shows a histogram of the FWHMs of the real GCs (secure sample)
compared with the FWHMs of the simulated objects (King profiles convolved with the
instrument PSF).
The excellent agreement between the two histograms, both at the ``peak'' and the
``tail'' of the distributions (the small discrepancies are most probably
related to the fact that only one PSF model has been used for the whole mosaic),
confirms that the simulated objects are representative
of a population of GCs spanning the whole range of shape parameters
(i.e., $r_{c}$ and $c$) expected for real sources.
Moreover, it ensures that the ``tail''
of the FWHM distribution is not due to contamination by spurious sources, but it is
produced by real GCs (and precisely by the most extended and faint). The ``tail''
contains a significant fraction of the sources; for example, the fraction of
GCs with FWHM between 3 and 4 pixels equals 10\% of the whole sample.
In $\S$\ref{Globular Cluster Selection}, we reported that
we imposed a FWHM limit of 4 pixels when selecting GCs.
Based on the results of the simulation, we remark that
choosing a more strict limit, although helping in excluding non-GC objects
(e.g. galaxies), would also significantly reduce the number of real GCs in
the candidate list.

We further tested the FWHM limits by simulating the most extreme faint and extended
GCs and measuring their size.
We defined the limiting magnitude for this simulation as that corresponding to
the 90\% completeness (calculated with the method described above,
but excluding the FWHM criterion).
We set the $c$ parameter to 2.0 and we chose a $r_c$ of
4~pc (0.26~pixels), which is in the upper range of observed
core radii $r_c$ \citep[see][]{jor05}.
The limiting FWHM found with this method was $\sim$4~pixels,
in agreement with the adopted selection criteria.

\section{Properties of the Globular Clusters}
\label{Properties of the Globular Clusters}

\subsection{$V$ band Luminosity Function}
\label{$V$ band Luminosity Function}

The GC populations of almost all galaxies show a remarkable characteristic:
their Luminosity Function (GCLF), plotted in magnitude units, is commonly
represented by a Gaussian peaking at $M_{V}\sim-7.4$~mag, with a dispersion of
$\sigma_{V}\sim1.2$~mag \citep[e.g.][]{har91,ash98}, although minor deviations from
this shape may be expected for a number of reasons \citep[e.g.][Villegas 2010
for an extensive discussion on Gaussian representation of the GCLF in early type
galaxies]{jor07}. A formation scenario that would justify a Gaussian distribution
does not exist.
Indeed, the GCLF has been tested against other types of distributions, e.g., the
student $t_{5}$ function \citep{sec92}.
For this reason, and given the skewness of the GCLF, the peak of the distribution
is usually referred to as ``turnover``.
The location of the peak appears to be fixed and therefore, when expressed
in terms of the apparent magnitude, it can be used to estimate the galaxy distance
\citep{kun01}.

The left panel of Figure \ref{V_histograms} shows the $V$ band histogram for
the candidate GCs (light gray), the secure sample of GCs (dashed shadow) and
contaminating stars (gray shadow).

The expected contamination due to foreground stars has been evaluated using the
{\footnotesize TRILEGAL} online tool\footnote{http://stev.oapd.inaf.it/cgi-bin/trilegal}
\citep{gir05}, which simulates the spatial distribution of stars in the Milky Way.
Assuming models for the star formation rate, age-metallicity relation and initial mass
function of the Galaxy, {\footnotesize TRILEGAL} computes the age, metallicity,
mass and apparent photometry for the expected population of Galactic stars along the
line of sight within the desired field of view.
We queried the tool using the default parameters (Chabrier function for initial mass
function, exponential disk model for extintion, thin disk + halo + bulge components
for galaxy structure), for HST/WFPC2 (Vega system) magnitudes down to
$m_{_{V}}$ = 28~mag and for a field equivalent to the size of the $D_{25}$
of NGC4261.
Inspecting the histogram (Figure \ref{V_histograms}) we can see that the
foreground star contamination is negligible. Moreover, this estimate should be
considered as a rigorous upper limit, since this population was not filtered according to
the selection procedure applied to the GCs ($\S$\ref{Globular Cluster Selection}).

\subsubsection{Fitting The Luminosity Function Turnover}
\label{Fitting The Luminosity Function Turnover}
\begin{figure*}
 \makebox[\linewidth]{
   \includegraphics[width=0.48\textwidth,angle=0]{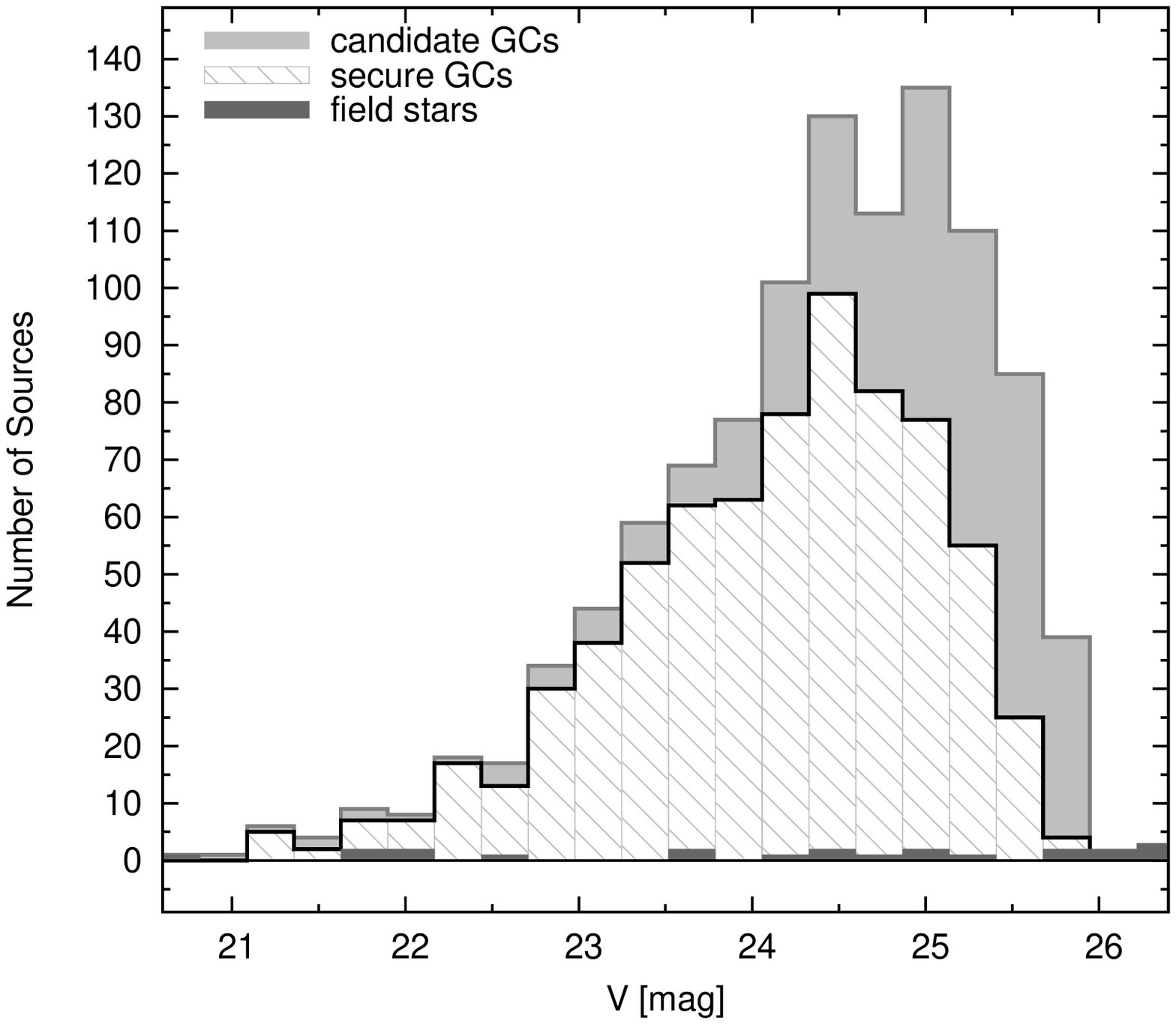}
   \includegraphics[width=0.48\textwidth,angle=0]{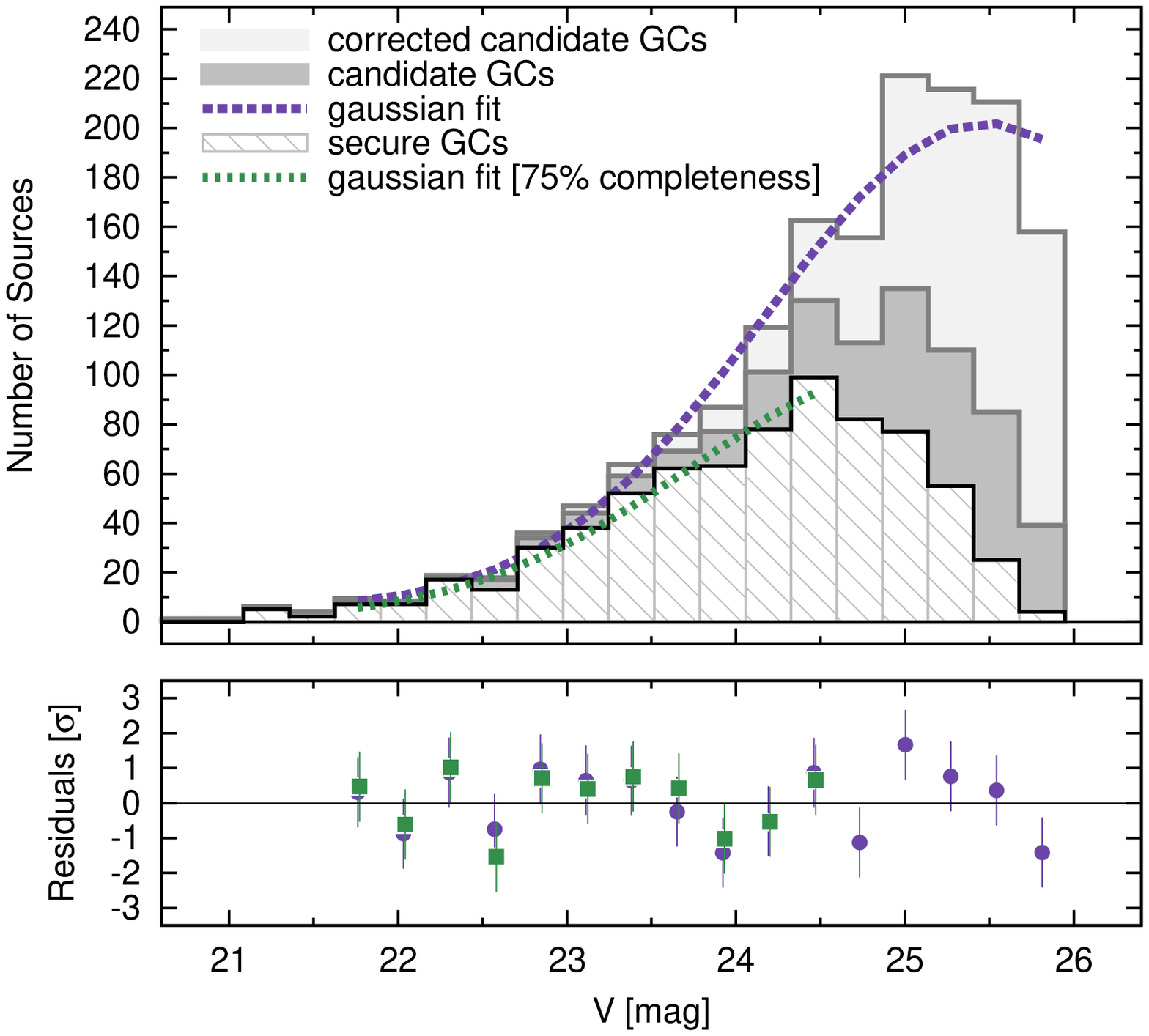}
  }
\caption{
 $V$ magnitude histograms for the GCs candidates and Gaussian fit.
 \emph{Left.} ($V$ magnitude histograms) - Candidate GCs are shown in light gray,
 the subsample of candidate GCs selected in colour (secure GCs) with a dashed shadow
 and the contaminating stars (from the {\tiny TRILEGAL} simulation) in gray shadow.
 \emph{Right.}
 (Incompleteness-corrected $V$ magnitude histogram and fits) - Candidate GCs are shown in
 light gray, the subsample of candidate GCs selected in colour (secure GCs) with a
 dashed shadow. The light gray area represents the distribution of the candidate
 GCs after applying the incompleteness correction, performed as described in
 $\S$\ref{Evaluating Incompleteness}. The fit refers to the secure sample (green dashed line)
 and to the corrected candidate sample (violet dashed line).
 The bin sizes have been optimized according to the Freedman-Diaconis criterion
 \citep{fre81}, using the mean photometric error as the minimum allowed value.
 \label{V_histograms}
}
\end{figure*}

We fitted a Gaussian profile to the histogram of the secure (colour selected) GC
sample (shown in Figure \ref{V_histograms}) in order to derive the peak location
and FWHM of the GCLF.

Notice that the distributions appear as skewed Gaussians due to the cutoff imposed
by the incompleteness at faint fluxes.
In particular, due to the colour selection and the more significant effect of the
incompleteness in the $I$ band data (caused by the higher background; see Figure
\ref{incompleteness}), the faint end of the  GCLF of the secure GC sample might be
biased by the high $I$ band detection threshold.
If this were to be the case, the result would appear as a deficit of
blue objects at faint magnitudes. Instead, comparing the colour histograms of objects
brighter and fainter than $m_{_{V}}>$24~mag, we find that such a selection effect
is minimal.

Nevertheless, in order to obtain a conservative result, we estimated the magnitude at
which such incompleteness starts affecting the sample using the following argument.
The faintest object detected in the $I$ band, at the 75\% completeness level,
has a magnitude $m_{_{I}}\sim24.0$~mag.
Therefore, according to the minimum colour of the blue GC subpopulation ($V-I\sim$0.6~mag;
see $\S$\ref{Globular Cluster Selection}),
a faint blue GC is expected to have a $V$
magnitude of $\sim$24.6~mag (which is close to the 75\% completeness level for
the $V$ band).
We adopted this value as the limit at which we ``truncated'' the fit of the GCLF.

The fit was performed using the \emph{Sherpa} package v4.2.1
(which is part of the {\footnotesize CIAO} tool suite v4.2).
The number of sources per bin of the Luminosity Function (LF) was high enough to
allow the use of $\chi^{2}$ statistics.
The uncertainties for each bin were estimated assuming a Poissonian distribution.
The errors we report refer to the
1-$\sigma$ (68.3\%) confidence bounds from the $\Delta\chi^{2}$ projection
for 1 interesting parameter.
The number of foreground stars and other
contaminating objects has been accounted for by including a constant in the fit.
The resulting value is compatible with the estimated number of contaminating stars
per bin (see Figure \ref{V_histograms}).

The turnover of the GCLF is located at $m_{_{V}}$ = 25.1$_{-0.6}^{+1.0}$~mag, and the
Gaussian distribution has a FWHM of 3.1$_{-0.8}^{+0.6}$~mag
(or $\sigma$ = 1.3$_{-0.3}^{+0.3}$~mag). Using the Third Reference Catalogue of
Bright Galaxies (RC3; \citealt{dev95}) we verified that reddening effects on peak
location are negligible ($\Delta_{B-V}$ = 0.02~mag).
The large errors on the peak location and FWHM are a consequence of the necessity
of truncating the fit around the turnover.
Nevertheless, the best fit FWHM value is in close agreement with
the results for the sample of elliptical galaxies studied by \cite{kun01} using
data obtained with a similar configuration (\emph{WFPC2}, filters F555W and F814W).
The results of the fits are shown in Figure \ref{V_histograms} and Table \ref{fit_V}.

The turnover of the GCLF has been proven to be a reliable distance indicator
\citep{kun01}.
Adopting a $M_{V}^{GC} = -7.4$~mag for the peak of the GCLF
(with an error negligible in comparison to our photometric errors)
we derived a distance modulus $m-M$ = 32.5$_{-0.6}^{+1.0}$~mag.
The corresponding distance is 31.6$_{-8.7}^{+14.6}$~Mpc,
in agreement with the estimate by \cite{jen03}
(29.4 $\pm$ 2.6~Mpc) based on surface brightness fluctuations.
\cite{tul88} measured a heliocentric velocity $v$ = 2202 $\pm$ 75~km/s for NGC~4261,
accounting for the infall of the Local Group to the Virgo cluster according to the
model described in \cite{tul84}.
This velocity corresponds to a distance of 31.3 $\pm$ 1.2~Mpc
\citep[adopting an $H_{_{0}}=70.4\pm1.4$~km/s/Mpc - Seven Years WMAP Observations;][]{jar11},
which is consistent with our measurement at the 1-$\sigma$ level.

\begin{table*}
 \centering
  \caption{G{\tiny CLF} F{\tiny IT} R{\tiny ESULTS}\label{fit_V}}
  \begin{tabular}{@{}cccccccc@{}}
   \hline
   Sample        & Peak                  & FWHM                 & Amplitude$^{\dagger}$ & Constant$^{\dagger}$ & Integrated Number     & Distance             & $S_{N}$     \\
                 & [mag]                 & [mag]                &                       &                      & of Sources            & [Mpc]                &             \\
   \hline
   Secure GCs    & 25.1$_{-0.6}^{+1.0}$  &  3.1$_{-0.8}^{+0.6}$ & 101$_{-21}^{+53}$     & 1.2$_{-1.2}^{+4.6}$  & 1242                  & 31.6$_{-8.7}^{+14.6}$ & 1.5$\pm$0.3 \\
   Candidate GCs & 25.5$_{-0.2}^{+0.2}$  &  3.0$_{-0.3}^{+0.4}$ & 196$_{-12}^{+14}$     & 5.2$_{-3.9}^{+3.2}$  & 2363                  & 38.0$_{-3.5}^{+3.5}$  & 2.8$\pm$0.5 \\
  \hline
  \end{tabular}
 \\

 \begin{flushleft}
 
  $\dagger$ Value of [central] bin. Bin size: $\Delta$$V$ = 0.27~mag. \\

 \end{flushleft}

\end{table*}

\subsubsection{GC Specific Frequency}
\label{GC Specific Frequency}

The GC Specific Frequency (SFs; number of GCs per unit luminosity) of a galaxy
is defined as $S_{N}=N_{GC}\times10^{0.4(M_{V}+15)}$ \citep[][]{har81}.

An estimate of the total number of GCs ($N_{GC}$) within the $D_{25}$
was determined for both the secure and the candidate GCs by fitting
their histograms  with a Gaussian and a constant which accounted for contaminating objects.
The LF for the candidate GCs sample had been previously corrected for incompleteness
effects by splitting the sample into the three galactocentric ranges defined in
Figure \ref{incompleteness}, applying the corresponding detection rate curve, and
then stacking back the results into one histogram, shown in Figure \ref{V_histograms}.
The parameters of the best fit Gaussian are shown in Table \ref{fit_V}.

Notice that the distance modulus derived from the fit of the LF for the candidate GCs
(see Table \ref{fit_V} and $\S$\ref{GC Specific Frequency}) is evidently
overestimated. The peak of the Gaussian is driven towards fainter
fluxes where the contamination by spurious sources is amplified by the
high completeness correction factor applied.
Unfortunately, these sources cannot be distinguished from real GCs using the
FWHM and elongation criteria.
For this reason, the total number of GCs resulting from the candidate sample
should be considered as un \emph{upper} limit. On the other hand, the number derived
from the secure sample, which was not corrected for incompleteness,
represents a \emph{lower} limit

The integral of the GCLF models yielded $N_{GC}\sim$1242 and $N_{GC}\sim$2363,
for the secure and candidate sample respectively.
Assuming an integrated magnitude for the galaxy of $m_{_{V}}$ = 10.01 $\pm$ 0.06~mag
(derived as described below) and a distance of 29.4 $\pm$ 2.6~Mpc \citep{jen03},
we calculated that: $S_{N} = $1.5 $\pm$ 0.3 for the secure sample and
$S_{N} = $2.8 $\pm$ 0.5 for the candidate sample. Both numbers are within the 
range of typical values for elliptical galaxies \citep[e.g.][]{kun01}.

The galaxy asymptotic magnitude was determined from a $2D$ fit performed on
the $V$ band WFPC2 data using {\footnotesize GALFIT} \citep{pen02}.
This software is able to fit the surface brightness distribution of a galaxy
to a $2D$ model accounting for the effect of $PSF$ blurring.
We chose a Sersic model
(plus a constant component accounting for the sky flux)
and used the SE{\footnotesize XTRACTOR} results for the galaxy
as initial fit parameters. We used a TinyTim $PSF$ image
(generated as described in $\S$\ref{Evaluating Incompleteness}),
while we let {\footnotesize GALFIT} produce the weight
image independently. We also applied a mask to exclude the central dust ring
from the fit. The best-fit model has a Sersic index of 5.7,
effective radius of $\sim$70$\arcsec$, major axis Position Angle (P.A.)
of -21~deg and axis ratio ($b/a$) of 0.8.
The ratio of the sum of residuals over the model-integrated light is less than
2\% for all the filters, indicating that the model is a good
representation of the starlight distribution of this galaxy.
The galaxy magnitude was derived from the integral counts under the model
applying the conversion formulae of \cite{hol_a}.

The model-subtracted image showed a residual light ring around
the galactocentric radius $R_{GC}\sim15\arcsec$, while we did not find evidence
for the tidal arm reported by \cite{tal09} (probably due to the lower exposure
time).
The peculiar pattern of the residuals is most likely related to the boxy
nature of the galaxy. {\footnotesize GALFIT} is able to fit a boxy model with a fixed
boxiness, while probably the isophotes of NGC~4261 have a radial
variation in this parameter \citep[such behavior has been reported in many
ellipticals; e.g.][]{nie89}.

\subsection{Colour Distribution}
\label{Colour Distribution}
An important result regarding GCs in massive galaxies is their bimodal colour
distribution \citep[e.g.][]{ash98,pen06}, generally attributed
to a metallicity difference between two subpopulations \citep[e.g.][]{bro06}.
There are different scenarios which can account for such bimodality.
The best supported is probably the merging scenario, according to which the
bluer (i.e., metal poorer) GCs are ``donated'' by merging galaxies, while the red (i.e. metal
richer) GCs are formed in the merger, along with the bulk of the galaxy field stars
\citep{ash92}. This model has the advantage of naturally explaining the
subpopulations and their different radial distributions (see Section
$\S$\ref{Spatial Distribution of Sources}), but does not address
the extent to which the newly formed GCs can account for the significantly
higher SF measured in elliptical galaxies with respect to those of the
spiral galaxies \citep[e.g.][]{har91}, which are supposed to be their progenitors.
It is also not yet clear whether a significant fraction of the young
star clusters formed in merging galaxies can survive to account for the red GCs population
of ellipticals.
Valid alternatives are the multi-phase dissipational collapse model \citep{for97},
which assumes that blue and red GCs are formed at different stages of the galaxy formation,
and the accretion scenario \citep{cot98}, in which the red GCs represent the intrinsic GC population
of the galaxy while the blue GCs are acquired from lower mass galaxies during dry mergers
or by tidal stripping.
As reviewed by \cite{bro06}, the three models are not mutually exclusive, and their
effects may sum to generate the observed GC subpopulations in ellipticals.

\begin{figure}
 \includegraphics[width=0.48\textwidth,angle=0]{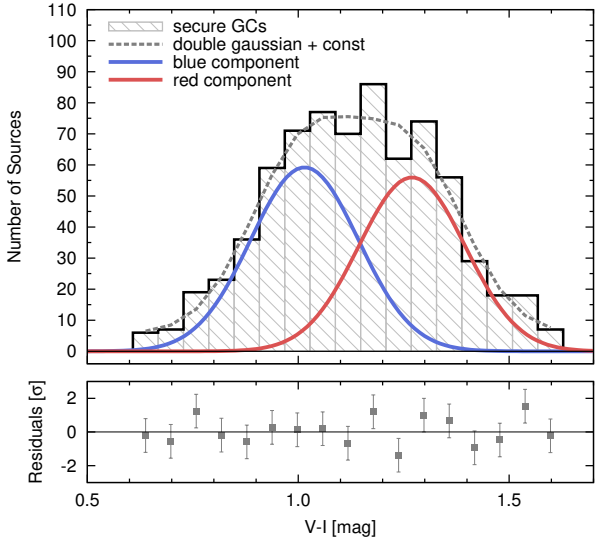}
 \caption{ 
  $V$-$I$ colour histogram for the secure GCs sample and fit.
  The data have been fit using two  Gaussians plus a small constant used
  to account for contamination.
  In the figure, we show the two Gaussian components (blue and red dotted lines)
  and the total convolved model (gray dashed line).
  The additive constant is not showed in the figure. 
  The bin sizes have been optimized according to the
  Freedman-Diaconis criterion \citep{fre81},
  using the mean photometric error as the minimum allowed value.
  \label{VI_histograms}
 }
\end{figure}

Figure \ref{VI_histograms} shows the $V$-$I$ histogram for the secure GCs sample.
The GC colour distribution doesn't show the clear bimodality typical of the GC
colour distribution in early-type galaxies \citep[e.g.][]{bro06}.
In order to quantitatively discern the number of components in the distribution,
we performed a mixture model analysis using the NMIX method of \cite{ric97}, which
implements a Bayesian test of univariate normal mixtures.
We run the code following the suggestions reported in \cite{ric97} about the priors
and hyperpriors (namely, choosing $\xi$ = 1.1, $K$ = 2, $h$ = 2, and setting the rest as default),
even though the test was demonstrated to be robust with respect to the choice of the initial
parameters. The test did not show significant evidence for a bimodal GC colour
(Bayes factor of 1 and 0.5, for the unimodal and bimodal models respectively).
However, the range of colours of our GC sample spans the typical colours
of both the blue and red GCs in elliptical galaxies, while it is too
wide to arise from a single GC population, suggesting that most probably the two
subpopulations are blended in our data. Moreover, as pointed out by \cite{pen06} in
their work on early-type galaxies in the Virgo cluster, even for those galaxies in which the
relatively low number of GCs does not allow one to distinguish between unimodal and bimodal
colour distributions, the result of a decomposition into two components provides a
reliable description of the system.
For these reasons, we assume that the distribution consists of superimposed blue and red
subpopulations and so we fit the colour histogram using two Gaussian
components, plus a constant to account for the contamination by
foreground stars (estimated as 1-2 objects per bin using the {\footnotesize TRILEGAL}
results - see $\S$\ref{$V$ band Luminosity Function}) and background galaxies.
The uncertainties for each bin were estimated assuming a Poissonian distribution.

Table \ref{fit_VI} reports the result of the fit.
For completeness, the table also shows the result from a single Gaussian fit.
In the double-component case, the large errors in the Gaussian normalizations
reflect the fact that the two amplitudes are strongly correlated.
Nevertheless, the locations of the peaks are accurate enough to allow us to identify a
blue and a red subpopulation, with average colours $V$-$I$=1.01$_{-0.06}^{+0.06}$~mag
and $V$-$I$=1.27$_{-0.08}^{+0.06}$~mag respectively. These values are in good agreement
(within 1$\sigma$) with those reported by \cite{lar01} based on a similar
analysis of the GC population of a sample of nearby early-type galaxies,
observed using deep exposures in the F555W ($V$) and F814W ($I$) \emph{HST} filters.

The best-fit model along with the individual components are shown in the top
panel of Figure \ref{VI_histograms}. The bottom panel shows the residuals with respect
to the best-fit model in units of standard deviation.

\begin{table}
 \centering
  \caption{$V$-$I$ F{\tiny IT} R{\tiny ESULTS}\label{fit_VI}}
  \begin{tabular}{@{}lccc@{}}
   \hline
   \multicolumn{4}{c}{\tiny{DOUBLE COMPONENT}} \\
   \hline
   Component     & Amplitude$^{\dagger}$ & Peak                    & FWHM                    \\
                 &                       & [mag]                   & [mag]                   \\
   \hline
   Blue Gaussian & 59$_{-29}^{+23}$      & 1.01$_{-0.06}^{+0.06}$  & 0.30$_{-0.06}^{+0.08}$  \\
   \addlinespace 
   Red Gaussian  & 56$_{-22}^{+15}$      & 1.27$_{-0.08}^{+0.06}$  & 0.30$_{-0.07}^{+0.1}$   \\
   \addlinespace 
   Constant      & 6$_{-3}^{+2}$         & -                       & -                       \\
   \hline
   \multicolumn{4}{l}{$\chi_{\nu}^{2}$ = 1.11, 10 d.o.f. ($P_{(Q-value)}$ = 35\%)}           \\
   \\[1pt]
   \hline
   \multicolumn{4}{c}{\tiny{SINGLE COMPONENT}} \\
   \hline
   Component     & Amplitude$^{\dagger}$ & Peak                    & FWHM                    \\
                 &                       & [mag]                   & [mag]                   \\
   \hline
   Gaussian      & 81$_{-4}^{+4}$        & 1.14$_{-0.01}^{+0.01}$  & 0.49$_{-0.04}^{+0.	02}$ \\
   \hline
   \multicolumn{4}{l}{$\chi_{\nu}^{2}$ = 1.06, 14 d.o.f. ($P_{(Q-value)}$ = 39\%)}           \\
  \end{tabular}

   \bigskip
   $\dagger$ Value of [central] bin. Bin size: $\Delta$$V-I$ = 0.06~mag. \\

\end{table}

We arbitrarly divided the secure GCs into blue and red by setting the threshold at
$V$-$I=$~1.15.
The discriminating value is representative of the 50\% probability for a GC
to belong to the blue or red subpopulation (the intersection of the Gaussians
in Figure \ref{VI_histograms}). This value is consistent with the separation of the
two subpopulations in other galaxies \citep[e.g.][]{ash98}.

In order to estimate the relative abundance of the blue and red subpopulations, we
performed the following Monte-Carlo simulation.
We sampled each of the parameters of the two-Gaussian fit (described above).
The sampling distribution was a multi-variate Gaussian with a mean equal to the best-fit value
of each parameter and a FWHM derived from the covariance matrix of the fit.
At each trial, the set of sampled parameters defined two Gaussians whose flux
fractions with respect to the total flux from the combined model were recorded.
Finally, the expectation value for the fraction of the blue/red component
was calculated as the median over all the trials.
With this method, we estimated that the blue (red) GCs represent the
60$_{-24}^{+21}$\% (40$_{-21}^{+24}$\%) of the total sample of 718 GCs.
The upper and lower uncertainties on the fraction represent the 68.3\%
quantile (1 $\sigma$) around the medians.

\section{Spatial Distribution of Sources}
\label{Spatial Distribution of Sources}

Prompted by the findings of \cite{zez03} and \cite{gio05}, we investigated the spatial
distribution of the secure GC sample.

The region of the galaxy within $R_{GC} <$ 25\arcsec~was excluded from the analysis
due to the high background which drastically limited the source detection sensitivity
(see $\S$\ref{Evaluating Incompleteness}).
Since the GC detection rate depends on the galactocentric distance
(Figure \ref{incompleteness}) and affects the relative number of detected sources at
each radius, introducing artificial asymmetries, we decided to limit the analysis to
those GCs having a magnitude brighter than the 75\% completeness limit for the $V$ band
(corresponding to $m_{V}$ = 24.6~mag).

\subsection{Radial Distribution}
\label{Radial Distribution}

\begin{figure}
 \includegraphics[width=0.48\textwidth,angle=0]{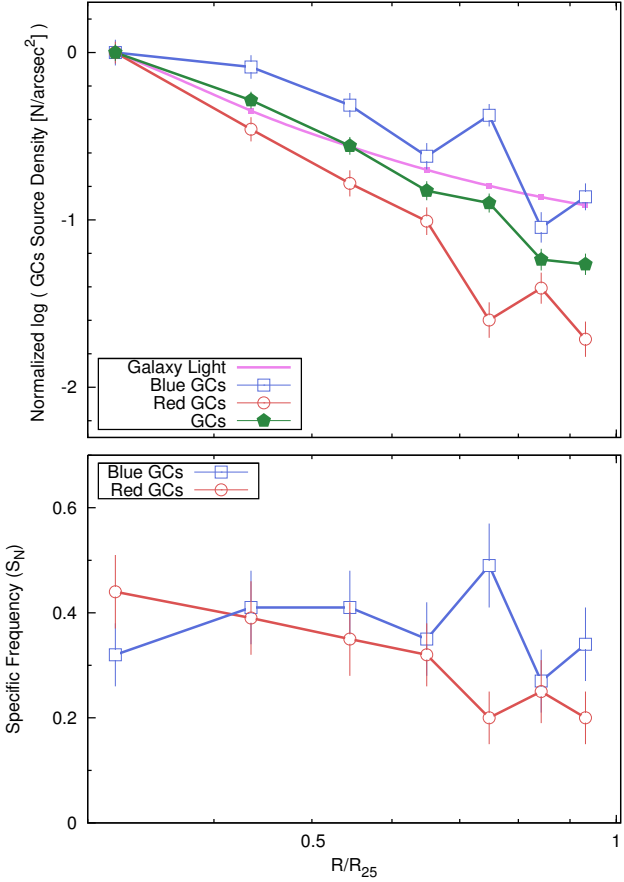}
 \caption{
  GCs radial profiles.
  \emph{Top.}
  Comparison of the source densities of the GCs sample (defined in Section
  $\S$\ref{Spatial Distribution of Sources}) and of the blue and red subpopulations
  (green, blue and red lines respectively) with the scaled light profile model of the
  galaxy (violet line).
  The data points represent the source densities evaluated over elliptical annulii.
  The radial coordinate is expressed in terms of the galaxy semi-major axis
  $R_{25}$ = 2.04\arcsec.
  The vertical error bars represent Poissonian errors.
  The profiles have been normalized in order to match at the radius of the first annulus.
  \emph{Bottom.}
  Radial profile of the Specific Frequency ($S_{N}$) of the blue and red subpopulations
  (blue and red lines respectively), calculated as
  $S_{N}(i)=N_{GC}^{i}\times10^{0.04(M_{V}^{i}+15)}$, where $i$ is the annulus index.
  \label{radial}
 }
\end{figure}

In the top panel of Figure \ref{radial} we present the radial profiles of the blue and
red GCs. As a reference, we also plot a scaled $V$-band light profile model of NGC~4261
(as derived from the {\footnotesize GALFIT} fit; see $\S$\ref{GC Specific Frequency}).
The data points represent the surface densities (number of sources per arcsec$^{2}$)
over elliptical annulii centered on the galaxy center. The orientation and ellipticity
of the annulii have been chosen to match the P.A. and ellipticity of the galaxy light
profile.
The profiles have been normalized to match at the first point.

We used \emph{funtools} \citep{man01} to count the number of sources within each annulus.
The annulii sizes (i.e. binning of the plot) have been chosen in order to contain
a statistically significant number of sources (at least 20 sources per bin)
that would allow the use of a $\chi^{2}$ test.
The uncertainty for each bin is simply the square root of the counts divided by
the annulus area.

The distribution of the overall GC population looks consistent with the trend
of the light profile, at least from the galaxy center up to $R/R_{25}$ = 0.75,
where it starts declining, mostly driven by the drop of the red GC component.
The difference between the trends of the blue and the red GCs radial profiles
is more evident when the distributions are plotted in terms of Specific Frequency
($S_{N}$), calculated as $S_{N}(i)=N_{GC}^{i}\times10^{0.4(M_{V}^{i}+15)}$
, where $i$ is the annulus index (bottom panel of Figure \ref{radial}).
In fact, recalling that $S_{N}$ expresses the relation between the number of clusters
and the galaxy light, the systematic decrease of $S_{N}^{_{RED}}$
denotes a clear inconsistency with the galaxy surface brightness.
On the other hand, $S_{N}^{_{BLUE}}$ varies around a constant value.

Such behaviour seems to be in contrast with the commonly accepted scenario
(see $\S$\ref{Colour Distribution}) according to which the red (young) GCs
follow the galaxy light, while the blue (old) are spread almost uniformly
(e.g. \citealt{bro06,ash98}).

On the other hand, the strong blending of the colours of the two populations
of our sample (see $\S$\ref{Colour Distribution}) may not have allowed us to
separate properly the two profiles.

A Kolmogorov-Smirnov (K-S) test has been used to compare the radial distribution
of blue and red GCs against a Sersic brightness profile
(whose parameters have been derived from the {\footnotesize GALFIT} fit;
see $\S$\ref{GC Specific Frequency}).
Since the Sersic profile is defined assuming circular symmetry, the dependence of the
galactocentric distance of a GC ($R_{GC}$) on the ellipticity of the galaxy had to
be removed as well.
The variable used to perform the test was the major axis $a$ of the
ellipse intercepting the GC and defined by the same center, elongation and P.A.
as for the galaxy. This can be calculated as:
 \begin{equation}
  a_{i} = { {R_{GC,i} \times \sqrt{e^{2}\cos^{2}(\theta_{i}-P.A.) + \sin^{2}(\theta_{i}-P.A.)} }\over{e} }
 \end{equation}
where $e$ is the elongation, $\theta$ the azimuthal angle ($\theta$ = 0
towards North and increases counterclockwise) and $i$ the GC index.
The test returned a probability well below the 0.1\% level for the blue GCs
and a probability of 0.3\% for the red GCs,
therefore rejecting the null hypothesis that their radial distributions
follow the galaxy light profile.

\subsection{Azimuthal Distribution}
\label{Azimuthal Distribution}
We evaluated the azimuthal distribution of the overall GC sample as well as of
the blue and red subpopulations by comparing their source densities against the
galaxy surface brightness within a set of elliptical wedges centered on the galaxy
nucleus. The parameters of the ellipses (center, ellipticity) have been obtained
from the {\footnotesize GALFIT} fit (see $\S$\ref{GC Specific Frequency}).
The angular extent of the wedges has been chosen to ensure Gaussian statistics
for the majority of the bins ($>$20 sources per bin).

In our mosaics, 4 areas within the galaxy $D_{25}$ were covered by the
WFPC2 PC camera (field of view of 35$\arcsec$$\times$35$\arcsec$).
We had to exclude these locations from the current analysis due to their different
sensitivity (which would influence the surface brightness estimation)
and higher noise (which could create artificial deficits of GCs).
When measuring the galaxy light, we also masked out all the GCs (secure sample),
foreground stars and background galaxies.

The measurements were repeated for different rotations (offsets) of the
ellipse wedges in order to address the significance of any density excess.
The wedge offset is defined on the rotation of a complete set of wedges measured
counterclockwise with respect to the North.
For each offset, a $\chi^{2}$ test between the azimuthal profiles of the GC density
and that of the galaxy light has been performed in order to identify the wedge rotation
for which the anisotropy is maximized.
Notice that, in this context, the azimuthal surface brightness distribution of
the galaxy simply represents a uniform comparison distribution which also includes
the modulation due to the geometry of the wedges alone.

\begin{figure}
 \includegraphics[width=0.48\textwidth,angle=0]{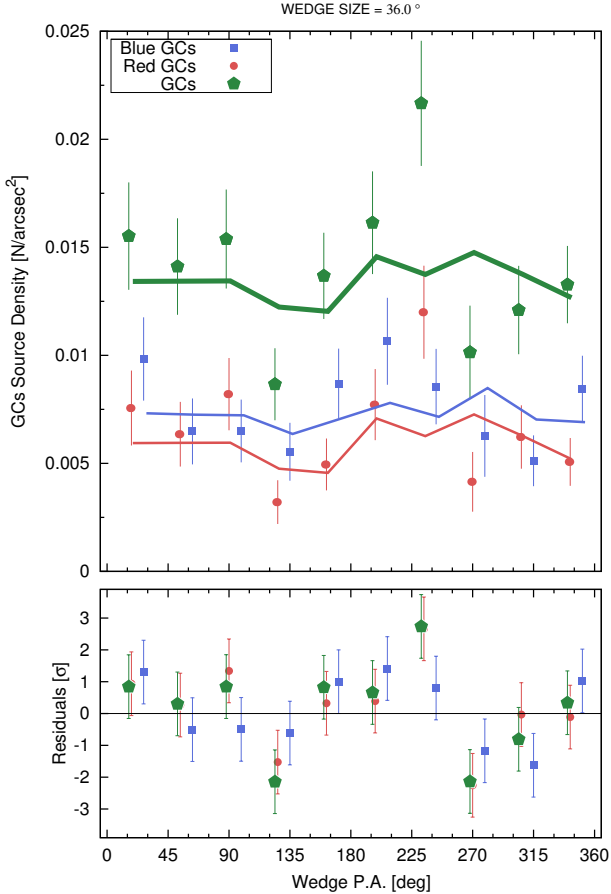}
 \caption{
  Source density azimuthal profile. The data points represent the density of GCs
  within wedges of 36$^{\circ}$ for the GCs sample (green diamonds) and for the red
  and blue subpopulation (red dots and blue squares respectively) as a function of
  the wedge P.A., for the wedges offset maximizing the anisotropy of each series
  (P.A. = 0 towards North and increases counterclockwise).
  The solid lines represent the best fit of the galaxy light to each data series.
  The bottom panel of the figure reports the scatter of the three data series
  with respect to the fit in units of $\sigma$.
  The data points have been slightly shifted along the $x-axis$ for presentation
  purposes.
  \label{pie}
 }
\end{figure}

Figure \ref{pie} shows the source density azimuthal profile of the whole GC sample
as well as the blue and red subpopulations for the wedge-offset that maximizes
the $\chi^{2}$ statistic for each series. The solid lines represents the best-fit
of the galaxy light to each data series (used to calculate the $\chi^{2}$).
The counter-intuitive modulation of the azimuthal distribution of the galaxy
light is due to a combination of two effects: the boxiness of
the galaxy (so that the wedges along the major axis have larger areas but
include less flux), and the masking we applied.
The bottom panel of the figure reports the fit residuals expressed in terms of $\sigma$
with respect to the fitted model (i.e. starlight distribution).

The plot shows a clear density enhancement (more than 2 $\sigma$) 
at P.A. $\sim$225$\pm$18$^{\circ}$, and significant deficits at P.A. $\sim$120$\pm$18$^{\circ}$
and at P.A. $\sim$270$\pm$18$^{\circ}$. The peak-to-peak variation is of the order of several
$\sigma$; the red GC subpopulation seems to follow the same modulation, while the
statistics for the blue distribution are too poor to derive any conclusion.

The GCs in the overdensity regions do not show any difference in brightness or colour distribution
with respect to the ``field'' GCs (the KS test can not distinguish the two
distributions), their $V$ and $V-I$ histograms closely resembling the ones of the
whole secure GCs sample (shown in Figures \ref{VI_histograms} and \ref{V_histograms}).

This test showed that the reduced $\chi^{2}$ ($\chi_{\nu}^{2}$) for the whole
sample was as high as 2.5 for 9 degrees of freedom.
Since we deliberately select the highest $\chi_{\nu}^{2}$, the usual
$\chi_{\nu}^{2}$ probability density function will not provide a correct
confidence level. Instead, we had to compare this value against the $\chi_{\nu}^{2}$
generated by a population of GCs distributed according to the galaxy light, in order
to exclude the possibility that the overdensities of GCs are statistical fluctuations.
To reproduce the proper $\chi_{\nu}^{2}$ probability distribution, we performed
the following test. We generated $n$ points (where $n$ equals the number of wedges)
distributed according to a Poissonian density function around the average number
of GCs found within a wedge. Such a set of data represented the equivalent of the
GCs azimuthal profile measured for a certain wedge rotation. The data were
then fit with a constant. The correct way to run this test would
have been to use the galaxy light azimuthal distribution to derive the average
of the Poissonian distribution in each wedge, but, since we assume a constant
$N_{GC}$ to light ratio in each wedge, the two methods are exactly equivalent
(in the latter case the appropriate fit would just be a rigid shift of that
curve along the y-axis).
We simulated a set for each wedge offset and we recorded the maximum $\chi_{\nu}^{2}$ among
the sets, as for the real data.

\begin{figure}
 \includegraphics[width=0.48\textwidth,angle=0]{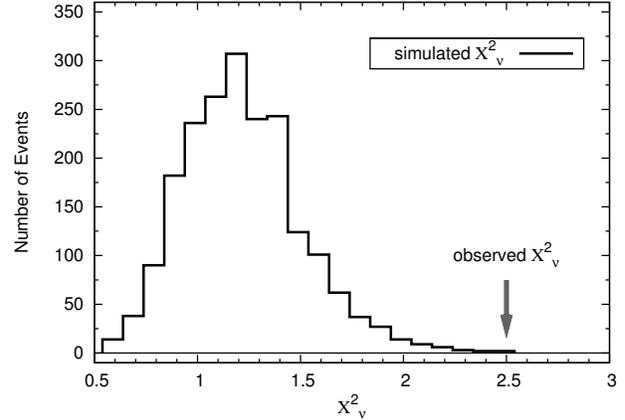}
 \caption{
  Probability distribution of $\chi_{\nu}^{2}$ produced by the fit of a constant to
  random sets of 10 data points distributed according to a Poisson distribution.
  The simulation is meant to reproduce the $\chi_{\nu}^{2}$ derived
  for a uniform azimuthal distribution (see text for more details). The arrow
  represents the $\chi_{\nu}^{2}$ obtained fitting the azimuthal distribution of
  the GCs to the galaxy light. The area covered by $\chi_{\nu}^{2}$ equal or higher
  than this value corresponds to a probability
  \mbox{$P_{\chi_{\nu}^{2}>\chi_{\nu, obs}^{2}}$~$<$~0.1\%}, therefore indicating
  that the asymmetry in the azimuthal distribution of the GCs with respect to the
  galaxy light is statistically significant.
 \label{pie_test}
 }
\end{figure}

The whole simulation was repeated 2000 times in order to generate the
$\chi_{\nu}^{2}$ probability distribution. The histogram in Figure \ref{pie_test} shows
the resulting distribution of the simulated max $\chi_{\nu}^{2}$.
The arrow shows the value of the $\chi_{\nu}^{2}$ measured from the data.
The integral of the simulated $\chi_{\nu}^{2}$ exceeding this value corresponds
to a probability \mbox{$P_{\chi_{\nu}^{2}>\chi_{\nu, obs}^{2}}$~$<$~0.1\%}.
This result indicates that the asymmetry in the azimuthal distribution of the GCs
with respect to the galaxy light is statistically significant at the 99.9\% level.

As a further test, we compared the distributions of the candidate GCs, and the blue
and red subpopulations, against a uniform distribution using a K-S test.
The KS test results did not allow us to reject the null hypothesis (i.e., that
the azimuthal profile follows a uniform distribution), for either of the three GCs
subpopulations (we obtained probabilities of 7\%, 35\% and 73\% for the whole
sample and the blue and red subpopulations, respectively).
In order to assess the significance of this result, we calibrated the K-S test in
a similar fashion as for the $\chi^{2}$ method.
For each population (secure GCs, red subpopulation and blue subpopulation),
we generated $m$ random angles (where $m$ is the number of sources in the population)
uniformly distributed in the range $[0,2\pi]$. The cumulative distribution of this
sample was compared against the cumulative of a uniform distribution and the
maximum distance $D$ between the two was recorded (such distance \emph{is} in fact
the K-S statistic).
The simulation was repeated 10000 times in order to generate the probability
distribution for $D$.
The integral of the values exceeding the $D$ measured in the K-S test on the
real data correspond to the probabilities $P_{D>D_{obs}}^{GCs}\sim10\%$,
$P_{D>D_{obs}}^{red}\sim70\%$ and $P_{D>D_{obs}}^{blue}\sim30\%$ for the secure GCs
sample and the red and blue subpopulations, respectively.
This result confirmed that the K-S test is not sensitive enough to identify the
differences seen between the samples.

\subsection{The two-Dimensional Distribution of the GC Population}
\label{The two-Dimensional Distribution of the GC Population}

\begin{figure*}
 \makebox[\linewidth]{
   \includegraphics[width=0.48\textwidth,angle=0]{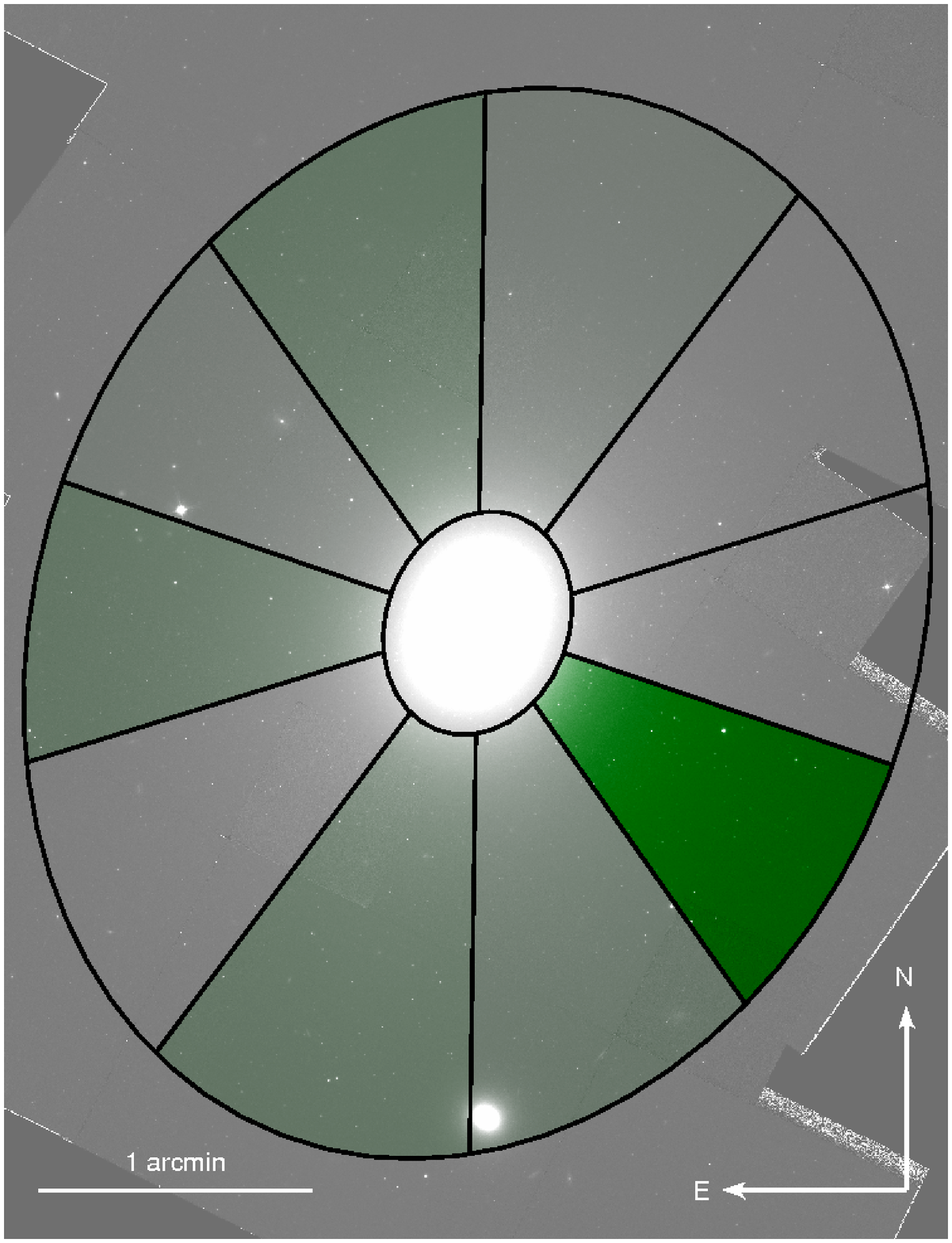}
   \includegraphics[width=0.48\textwidth,angle=0]{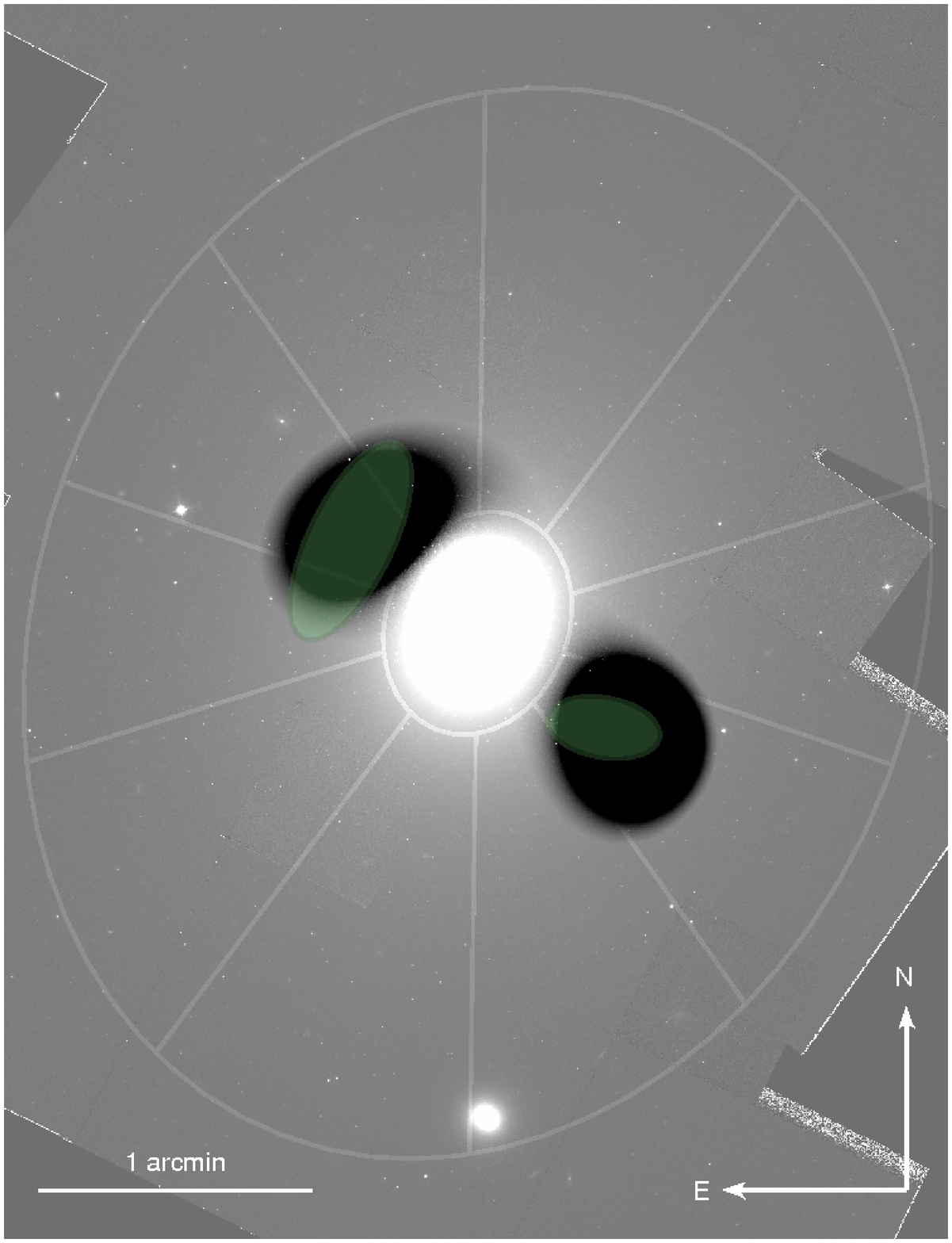}
 }
 \caption{
  Local enhancements of GCs.
  \emph{Left.}  
  Over a mosaic of NGC~4261 ($I$ band), we overlay
  the sectors over which we evaluated the GC azimuthal densities shown in
  Figure \ref{pie} (black ellipse).
  The inner and outer radius of the ellipse correspond to
  $R_{GC}$ = 25$\arcsec$ and $R_{GC}$ = $R_{25}$ respectively.
  The elliptical annular wedges we show are rotated to the angles at which the
  anisotropies in the GC azimuthal distribution are maximized. The filled wedges
  map the density enhancements shown in Figure \ref{pie}: the color intensity is
  proportional to the significance of the excess
  (i.e. value of the corresponding - positive - residual in bottom panel of Figure \ref{pie}).
  \emph{Right.}
  Overlaid on a mosaic of NGC~4261 ($I$ band), the shaded ellipses
  show local enhancements in the distribution of GCs detected with a Voronoi
  Tesselation and Percolation (VTP) algorithm.
  Each clustering is represented
  by an ellipse whose axes size equals the $\sigma$ of the spatial distribution of
  its GCs. The black shadows represent the areas of higher clustering of the
  secure GC sample as detected by the {\tiny CIAO} task \emph{csmooth}.
  The results from the VTP algorithm and from \emph{csmooth} are clearly consistent with
  those from the wedges test, and they point to a GC density excess along a NE-SW direction.
  \label{enhancements}
 }
\end{figure*}

In order to further investigate any local enhancements in the spatial
distribution of the GCs we used the {\footnotesize CIAO}\footnote
{
 The \emph{Chandra} data analysis package, supported by the Smithsonian Astrophysical Observatory.
}
tools \emph{csmooth} and \emph{vtpdetect}.

These tools were originally designed to identify clusters of photons in X-ray
images. For this reason, we created ``images'' of the GC locations to be used
by the tasks, with pixel value equal to 1 where a GC was present and 0 elsewhere.

The \emph{csmooth} tool smooths the image using an adaptive Gaussian kernel:
the scale for each area is increased until the counts under the smoothing kernel
exceed the significance threshold over the (local) background set by the user.
We set a minimal significance threshold of 3~$\sigma$. The background, which in
our case represents the uniform component of the GCs distribution, has been
estimated locally. The task is robust with respect to the
modification of the other parameters (such as the initial guess of the smoothing scale),
which mostly affected the computational time.

The result of this procedure is shown in Figure \ref{enhancements}.
We can clearly identify two regions where the clustering of GCs appears to be
significant (the regions appear slightly curved since we excluded sources near the
center of the galaxy).
These enhancements in the source density are located around P.A.$\sim$30$^{\circ}$
and P.A.$\sim$245$^{\circ}$ (NE and SW regions), in agreement
with the locations of the peak in the azimuthal distribution of GCs (Figure \ref{pie}).

\medskip

The \emph{vtpdetect} task exploits the Voronoi Tessellation and Percolation (VTP)
technique to detect connected sets of photons above a threshold density.
In our case, each GC played the role of a distinct photon. The task ``cuts out''
a triangular area around each source and calculates the source density (number/area).
The background model is estimated by fitting the low-end of the cumulative
density function (CDF). A density cutoff (used to distinguish areas containing
a source photon from background) is derived from the residual between the CDF
and the background model. Contiguous areas over the density cutoff are then
clustered together using a percolation algorithm. A source built in this way is
considered significant if it exceeds the desired false source limit due to
background fluctuations.
A detailed description of the VTP theory is presented in the
{\footnotesize CIAO} Detect Reference Manual (December 2006)\footnote
{
 http://cxc.harvard.edu/ciao/download/doc/detect\_manual/
 vtp\_theory.html
}.

In order to use \emph{vtpdetect} on non X-ray data we had to deduce the task
parameters through a trial-and-error approach.
We estimated a GC ``flux cutoff'' comparing the source density in the outer
regions of the image against the average area per source. The result was compatible
with the default value for the relevant task parameter (\emph{maxbkgflux} = 0.8).
We adopted the default values also for the other parameters related to the
background estimation (\emph{mintotflux} = 0.8, \emph{maxtotflux} = 2.6,
\emph{mincutoff} = 1.2, \emph{maxcutoff} = 3), after verifying that they did
not significantly affect the detection results. We set \emph{limit} = 0.001 and
\emph{scale} = 1; these values have been chosen in order to limit the
source deblending.

The 1$\sigma$ VTP ellipses defining source clustering are shown in
Figure \ref{enhancements}. The results from the VTP test confirm the excess
detected by \emph{csmooth}.

\section{Discussion}
\label{Discussion}

Using $HST$ WFPC2 data in the $B$, $V$ and $I$ band
(HST filters F450W, F606W and F814W, respectively) we characterized the GC population
of the nearby elliptical galaxy NGC~4261.
We studied the Globular Cluster Luminosity Function (GCLF), deriving a total number of
$\sim$2363 GCs, corresponding to a Specific Frequency (SF; number of GCs per unit
magnitude) of 2.8$\pm$0.5. From the location of the peak of the GCLF, we derived a distance
of 31.6$_{-8.7}^{+14.6}$~Mpc (assuming a $M_{V}\sim-7.4$~mag - e.g. \citealt{har91,ash98}).
The colour distribution can be interpreted as the superposition of a red and a blue
subpopulation, with average colour $V-I$ = 1.01$_{-0.06}^{+0.06}$~mag and
1.27$_{-0.08}^{+0.06}$~mag respectively.

Most importantly, we used different techniques to study the two-dimensional
distribution of the GCs sample and its blue and red subpopulations.
As discussed in $\S$\ref{Azimuthal Distribution} and
$\S$\ref{The two-Dimensional Distribution of the GC Population}, we discovered that the
GC population of NGC~4261 shows evidence for an asymmetry in the azimuthal distribution
above the 99.9\% confidence level.
A similar effect is noticed in the red GC subpopulation, but on
at lower significance. The result for the blue GCs is not statistically significant.
The origin for such anisotropy may reside in the evolution
of the galaxy in the recent past. Next, we discuss possible scenarios that address
to this asymmetry.

\begin{enumerate}
\item \emph{\textbf{Major merging event}}.
 The effects of a major merging event represent the most obvious candidate to
 explain such a significant asymmetry.
 NGC~4261 does not show signs of recent strong interactions (shells, ripples, arcs, tidal
 tails or other distortions), therefore we exclude a major merger
 within the last 1~Gyr, since this is the timescale on which
 interaction features fade away \citep[e.g.][]{qui84,her88,her89}.
 The difference between the starlight and GC distributions implies that, if the GC
 asymmetry is related to a major merger event (whether new GCs have been formed in situ,
 donated or just been displaced), the relaxation timescale for the GC population is
 larger than for stars. This is a complicated issue since GC systems extend beyond
 the main stellar body of the galaxy, and their dynamics at large galactocentric radii
 are affected by the dark matter (resulting in higher velocity dispersions for the GCs
 than for the stars). To our knowledge, there is no reported
 evidence for such a difference in the relaxation time so far.

 A major merging event may have triggered episodes of star formation within
 remnant tails along the line of sight. In this case, the enhanced GC density along
 the NE-SW axis of NGC~4261 may just be due to a projection effect.
 Studies of galaxies at an advanced state of merging, such as NGC~7252, NGC~3256 or
 NGC~3921 (\citealt{kni03,sch96}) show indeed that tidal tails host significant
 populations of proto-GCs.
 Therefore a certain fraction of GCs form in highly asymmetric distributions.
 The question is what fraction of these proto-GCs can survive tidal destruction
 and whether they can maintain their original spatial asymmetry during the subsequent
 dynamical evolution of the  host galaxy.

 In addition, there are cases suggesting incomplete violent relaxation which
 would naturally result in non uniform distribution.
 \cite{sch96} showed that the candidate GCs of the protoelliptical NGC~3921
 may have experienced the same (incomplete) violent relaxation as did the average
 starlight. Nevertheless, their data (although only partially covering the galaxy
 extent) show indications of asymmetry both in the radial and in the azimuthal
 distribution of the GCs. This suggests that spatial asymmetries of GC systems
 may indeed survive the relaxation process.

 Numerical simulations suggest that  massive boxy ellipticals
 (such as NGC~4261) are most probably the result of dissipationless (dry) mergers
 between ellipticals of similar size \citep[e.g.][]{kho05,naa06}.
 Since this scenario does not involve gas, or all the gas is converted to stars
 \emph{prior} to the collapse, the number of newly formed GCs along tidal tails
 would be negligible.

 Moreover, according to the merging scenario, the displaced GCs should mostly be
 younger (i.e. metal-rich). Since we observe asymmetry in both the blue and red
 subpopulations (although the result for the blue GCs is not statistically
 significant), we may conclude that the former hypothesis does not apply in
 the case of NGC~4261.

 Given the forementioned considerations, we conclude that a major merging event
 may be responsible for the observed asymmetry only if the relaxation timescale
 for the GC population could be proven to be larger than for stars.

\item \emph{\textbf{Minor merging event}}.
 As mentioned above, the absence of fine structure in NGC~4261 excludes the
 possibility of a recent ($\sim 1$~Gyr) major merging, but it does not rule out
 this possibility for minor mergers, since these events have weaker effects on
 the galaxy light distribution.
 This is the case in the elliptical galaxy NGC~1052, which underwent a recent
 (1~Gyr) merging event with a gas rich dwarf \citep{gor86}. The galaxy shows a weak
 rotation about its minor-axis (suggesting triaxiality) and significant minor-axis
 position angle swings \citep[see notes in][]{dav88}, but,
 similarly to NGC~4261, it exhibits a very low fine structure parameter \citep{sch92}.
 \cite{pie05} though, found no
 young GCs associated with the recent merger event of NGC~1052.
 This may indicate that no significant increase in the GC population should
 be expected even in case of gas rich minor merging.

 Non-relaxed GC distributions can be easily explained by GCs donated in
 a recent minor merging event but their number densities would be too small to
 generate statistically significant asymmetries over the galaxy light background.
 On the contrary, we can not rule out the possibility that the ``indigenous'' GCs
 have just been displaced during the merger.

\item \emph{\textbf{Galaxy interactions}}.
 Galaxy interactions are known to provoke displacements of the GCs systems.
 For example, the flyby encounter of NGC~1399 with the nearby NGC~1404
 disturbed the velocity structure of the outer regions of the GC system of
 NGC~1399 \citep{nap02}, or even generated a ``tidal stream'' of intergalactic
 GCs \citep{bek03}. A smoothed map of the displaced blue GCs candidates
 around NGC~1399 is reported in \citet[Figure 9]{bas06}.

 Moreover, simulations by \cite{ves00} showed
 that the perturbations on stellar systems of elliptical galaxies due to flyby
 encounters can last well past the closest passage of the perturber. Similarly, we can
 expect that the asymmetric distribution of the GC system of NGC~4261 may be
 a remnant from the recent interaction with one of its close by
 companions. In order to identify the candidate perturber, we calculated the
 RMS radial velocity of the members of the NGC4261 group \citep{gar93}
 and, considering a time between 1 and 2~Gyr for the remnants of the
 interaction to disappear, we estimated that the perturber should be found
 between 15$\arcmin$ and 30$\arcmin$ off NGC~4261.
 Unfortunately, even considering only the brightest galaxies ($g$ $<$ 14~mag)
 in this range, there are too many candidates to uniquely identify the
 possible perturber.

 In this sense, a close encounter similar to the one that happened between NGC~1399
 and NGC~1404 may have shifted some of the GCs of NGC~4261 from the NW-SE
 ``poles'' towards the NW-SE edge-on plane, with the galaxy nucleus preventing us from
 observing the distribution at the innermost radii.
 The GC systems extend much further than the host galaxy light, therefore the
 displacement may have affected only the outermost GCs without affecting
 the stellar orbits and hence the starlight. This event may also justify the peculiar
 drop of the GC radial profiles at large radii, shown in Figure \ref{radial}
 (although the effect may be due only to statistical fluctuations).
 If the GCs are indeed distributed continuously along a NW-SE axis,
 they would lie along a plane perpendicular to the rotation axis \citep{dav86},
 therefore the overdensity could be simply associated with the rotation of the galaxy.

 In addition, some GCs may have been stripped out of the interacting galaxy.
 However, since the candidate perturbers are significantly smaller than NGC~4261
 (therefore they have poorer GC systems),
 this effect can not explain alone the azimuthal variation of
 the GC density. In fact, its variation can reach up to $\sim$ 50\%
 (distance of the GC density from the fit of the galaxy surface brightness;
 Figure \ref{pie}), requiring a number of stripped GCs comparable to the system
 of NGC~4261.
 We find that there is no evidence for a difference between the properties (colour,
 FWHM, axis ratio, etc.) of the objects in the overdensity regions and
 those in the ``field''. This implies that the number of GCs subtracted from the
 interacting galaxy is relatively small or if the acquired GC population is
 numerically significant, then it has similar properties.

\end{enumerate}

The displacement of GCs under the effects of galaxy interaction or mergers and
estimation of the relaxation times for GC ``particles'' embedded in mixed
baryonic and dark matter environments is a subject that needs to be explored in
order to understand the distribution of the GC populations in the increasing
number of galaxies which show non-uniform GC systems.

\section{Conclusions}
\label{Conclusions}

We performed an analysis of the characteristics and spatial distribution of the
GC population of the nearby elliptical galaxy NGC~4261 using deep $HST$ WFPC2 data
in the $B$, $V$ and $I$ band (HST filters F450W, F606W and F814W, respectively).
Our photometry of the GC system resulted in a sample of
718 secure GCs, complete down to $m_{_{V}}$ = 23.8~mag (90\% completeness level).
We were able to derive the Globular Cluster Luminosity Function (GCLF) down to the
limiting magnitude $V$ = 24.6~mag.
The distribution peaks at $m_{_{V}}$ = 25.1$_{-0.6}^{+1.0}$~mag,
consistent with a distance of 31.6$_{-8.7}^{+14.6}$~Mpc (assuming a
peak $M_{V}\sim-7.4$~mag - e.g. \citealt{har91,ash98}).
We studied the colour distribution and probed the bimodal behaviour expected
in elliptical galaxies for GC systems \citep[e.g.][]{ash98}.
The $V-I$ colour distribution is compatible with the superposition of a blue
and a red component with average colour 1.01$_{-0.06}^{+0.06}$~mag and
1.27$_{-0.08}^{+0.06}$~mag respectively.
We calculated a Specific Frequency (number of GCs per unit
magnitude) of 2.8$\pm$0.5, therefore relating the GC abundance to the stellar mass.
We analyzed the two-dimensional distribution of the GC sample, and we found that
it shows evidence for an asymmetry in the azimuthal distribution above the 99.9\%
confidence level. A similar effect is noticed in the red GC subpopulation alone,
but at a lower level of statistical significance.
We discussed the origin of this anisotropy in context of galaxy
interaction and evolution.
Our main results regarding the GC spatial distribution are the following:

\begin{enumerate}
 \item The GC radial profile follows the galaxy surface brightness especially at
 galactocentric radii \mbox{$R/R_{25} <$ 0.6}.
 The red subpopulation profile is systematically steeper than the starlight, while the
 profile of the blue subpopulation looks more consistent with it (Figure \ref{radial}).
 \item The azimuthal distribution of the GC population shows significant
 enhancements along a NE-SW direction (Figure \ref{pie} and \ref{enhancements}).
 The red subpopulation seems to follow the same modulation, reflecting each local enhancement/deficit.
 The result for the blue GCs is statistically not significant.
 \item We suggest that the peculiar asymmetric distribution may be related to a past (dry)
 merger or interaction event. In particular, we favour the hypothesis that a fly-by encounter
 displaced the outermost GCs. The hypothesis of a past (older than 1~Gyr) major merging event
 cannot be ruled out, modulo the assumption that the GC relaxation time is longer than for
 the stars.
 We remark that the GCs may be distributed on an edge-on plane,
 since the galaxy nucleus prevents us from observing the distribution at the innermost radii.
 If this were to be the case, we cannot exclude that the overdensities could be
 rotation-related.
\end{enumerate}

The study of the anisotropy of the GC population of a galaxy is a relatively unexplored
field, which offers the unique opportunity of exploiting the GCs to investigate the
history of the host galaxy.
Spectroscopic data are critical to investigate whether the
GCs of NGC~4261 show any peculiar kinematics or they have typical dispersion velocities
for elliptical galaxies.
In particular, spectroscopy of the GCs in the overdensity regions will clarify
whether they are displaced GCs and move as an independent system from the overall GC
population.
Finally, we enphatize the need for numerical simulations aimed at modelling the
evolution of the GC spatial distribution after a galaxy interaction or a merger
and at estimating the relaxation times for GC ``particles''.

\section*{Acknowledgments}
We thank the anonymous referee for providing us with detailed and constructive
comments that have improved the quality of this manuscript.
PB and AZ acknowledge support by the EU IRG grant 224878.
Space Astrophysics at the University of Crete is supported by EU FP7-REGPOT grant 206469 (ASTROSPACE).
Support for the program GO-11339.01-A was provided through a grant from the Space Telescope Science Institute, which is operated by the Association of Universities for Research in Astronomy, Inc. under NASA contract NAS5-26555.
GT and AW acknowledge partial financial contribution from the ASI-INAF agreement I/009/10/0.
E.  O'Sullivan acknowledges the support of the European Community under the Marie Curie Research Training Network.
Funding for the Sloan Digital Sky Survey (SDSS) has been provided by the Alfred P. Sloan Foundation, the Participating Institutions, the National Aeronautics and Space Administration, the National Science Foundation, the U.S. Department of Energy, the Japanese Monbukagakusho, and the Max Planck Society. The SDSS Web site is http://www.sdss.org/.
The SDSS is managed by the Astrophysical Research Consortium (ARC) for the Participating Institutions. The Participating Institutions are The University of Chicago, Fermilab, the Institute for Advanced Study, the Japan Participation Group, The Johns Hopkins University, the Korean Scientist Group, Los Alamos National Laboratory, the Max-Planck-Institute for Astronomy (MPIA), the Max-Planck-Institute for Astrophysics (MPA), New Mexico State University, University of Pittsburgh, University of Portsmouth, Princeton University, the United States Naval Observatory, and the University of Washington.
This publication makes use of data products from the Two Micron All Sky Survey, which is a joint project of the University of Massachusetts and the Infrared Processing and Analysis Center/California Institute of Technology, funded by the National Aeronautics and Space Administration and the National Science Foundation.

\emph{Facilities:} HST(WFPC2)
\setcounter{table}{3}
\begin{landscape}
 \begin{table}
  \begin{minipage}{\linewidth}
   \caption{C{\tiny ATALOGUE} O{\tiny F} D{\tiny ETECTED} GCs\label{catalogue}}
   \begin{tabular}{@{}ccccccccccccc@{}}
    \hline
    GC          & RA (J2000)  & Dec (J2000) & $\in$       & FWHM        & Stellarity           & $\langle$S/N$\rangle$ & $B$         & $V$         & $I$          & $B$-$V$      & $V$-$I$ \\
    (ID)         & [hh:mm:ss] & [dd:mm:ss]  &             & [pixel]     &                      &                       & [mag]       & [mag]       & [mag]        & [mag]        & [mag] \\
    {\tiny (1)} & {\tiny (2)} & {\tiny (3)} & {\tiny (4)} & {\tiny (5)} & {\tiny (6)}          & {\tiny (7)}           & {\tiny (8)} & {\tiny (9)} & {\tiny (10)} & {\tiny (11)} & {\tiny (12)} \\
    \hline
  1 & 12:19:24.991 & +05:47:39.48 & 1.07 & 1.78 & 0.97 & 278.1 & 23.52$\pm$0.02 & 22.67$\pm$0.06 & 21.69$\pm$0.01 & 0.94$\pm$0.01 & 0.95$\pm$0.01\\
  2 & 12:19:22.278 & +05:47:40.68 & 1.15 & 1.68 & 0.94 & 32.4 & 26.13$\pm$0.07 & 25.28$\pm$0.06 & 24.08$\pm$0.02 & 0.94$\pm$0.05 & 1.25$\pm$0.04\\
  3 & 12:19:25.512 & +05:47:41.86 & 1.06 & 1.86 & 0.98 & 289.8 & 23.54$\pm$0.02 & 22.61$\pm$0.06 & 21.52$\pm$0.01 & 1.01$\pm$0.01 & 1.10$\pm$0.01\\
  4 & 12:19:24.360 & +05:47:42.57 & 1.12 & 2.15 & 1.00 & 89.5 & 24.87$\pm$0.03 & 24.13$\pm$0.06 & 23.34$\pm$0.01 & 0.84$\pm$0.02 & 0.67$\pm$0.02\\
  5 & 12:19:22.175 & +05:47:42.28 & 1.18 & 2.52 & 0.97 & 34.1 & 25.61$\pm$0.04 & 25.24$\pm$0.06 & 24.33$\pm$0.03 & 0.49$\pm$0.04 & 0.84$\pm$0.04\\
  6 & 12:19:26.772 & +05:47:44.50 & 1.31 & 2.05 & 0.97 & 42.4 & -           & 25.04$\pm$0.06 & 24.25$\pm$0.02 & -         & 0.68$\pm$0.04\\
  7 & 12:19:22.008 & +05:47:44.76 & 1.22 & 3.65 & 0.97 & 29.2 & -           & 25.43$\pm$0.06 & 24.44$\pm$0.03 & -         & 0.96$\pm$0.05\\
  8 & 12:19:24.579 & +05:47:45.71 & 1.03 & 1.91 & 0.95 & 103.5 & 25.00$\pm$0.03 & 23.95$\pm$0.06 & 22.75$\pm$0.01 & 1.12$\pm$0.02 & 1.25$\pm$0.02\\
  9 & 12:19:26.670 & +05:47:47.15 & 1.11 & 2.28 & 0.98 & 290.4 & 23.56$\pm$0.02 & 22.62$\pm$0.06 & 21.54$\pm$0.01 & 1.02$\pm$0.01 & 1.08$\pm$0.01\\
 10 & 12:19:22.306 & +05:47:46.92 & 1.07 & 1.83 & 0.97 & 243.8 & 23.68$\pm$0.02 & 22.82$\pm$0.06 & 21.75$\pm$0.01 & 0.95$\pm$0.01 & 1.07$\pm$0.01\\
 11 & 12:19:25.243 & +05:47:46.83 & 1.05 & 2.43 & 0.98 & 49.9 & 25.33$\pm$0.04 & 24.81$\pm$0.06 & 23.91$\pm$0.02 & 0.62$\pm$0.03 & 0.84$\pm$0.03\\
 12 & 12:19:25.279 & +05:47:50.12 & 1.29 & 1.93 & 0.98 & 557.4 & 22.45$\pm$0.02 & 21.58$\pm$0.06 & 20.47$\pm$0.01 & 0.96$\pm$0.01 & 1.13$\pm$0.01\\
 13 & 12:19:25.400 & +05:47:49.67 & 1.04 & 2.18 & 0.99 & 663.6 & 22.19$\pm$0.02 & 21.28$\pm$0.06 & 19.98$\pm$0.01 & 0.99$\pm$0.01 & 1.39$\pm$0.01\\
 14 & 12:19:27.281 & +05:47:49.17 & 1.36 & 3.03 & 1.00 & 123.3 & 24.76$\pm$0.03 & 23.77$\pm$0.06 & 22.80$\pm$0.01 & 1.08$\pm$0.02 & 0.92$\pm$0.01\\
 15 & 12:19:26.721 & +05:47:49.01 & 1.01 & 1.89 & 0.95 & 99.8 & 24.94$\pm$0.03 & 24.02$\pm$0.06 & 22.89$\pm$0.01 & 1.00$\pm$0.02 & 1.16$\pm$0.02\\
 16 & 12:19:24.447 & +05:47:51.17 & 1.15 & 2.22 & 1.00 & 158.7 & 24.37$\pm$0.03 & 23.41$\pm$0.06 & 22.49$\pm$0.01 & 1.05$\pm$0.01 & 0.85$\pm$0.01\\
 17 & 12:19:20.828 & +05:47:50.86 & 1.41 & 2.23 & 0.95 & 88.7 & 25.31$\pm$0.04 & 24.16$\pm$0.06 & 22.98$\pm$0.01 & 1.22$\pm$0.03 & 1.22$\pm$0.02\\
 18 & 12:19:21.749 & +05:47:52.19 & 1.12 & 1.89 & 0.97 & 181.0 & -           & 23.25$\pm$0.06 & 22.20$\pm$0.01 & -         & 1.04$\pm$0.01\\
 19 & 12:19:23.044 & +05:47:52.45 & 1.59 & 2.30 & 0.98 & 75.1 & 25.39$\pm$0.04 & 24.27$\pm$0.06 & 23.06$\pm$0.01 & 1.20$\pm$0.03 & 1.26$\pm$0.02\\
 20 & 12:19:23.052 & +05:47:53.10 & 1.97 & 1.89 & 0.81 & 124.6 & 24.63$\pm$0.03 & 23.67$\pm$0.06 & 22.80$\pm$0.01 & 1.05$\pm$0.02 & 0.78$\pm$0.01\\
 21 & 12:19:26.371 & +05:47:52.93 & 1.07 & 2.17 & 0.96 & 72.4 & 26.03$\pm$0.06 & 24.39$\pm$0.06 & 23.16$\pm$0.01 & 1.68$\pm$0.04 & 1.29$\pm$0.02\\
 22 & 12:19:26.542 & +05:47:53.69 & 1.25 & 2.44 & 1.00 & 147.1 & 24.37$\pm$0.03 & 23.53$\pm$0.06 & 22.45$\pm$0.01 & 0.93$\pm$0.01 & 1.09$\pm$0.01\\
 23 & 12:19:23.591 & +05:47:54.44 & 1.05 & 1.82 & 0.95 & 122.4 & 25.12$\pm$0.04 & 23.71$\pm$0.06 & 22.51$\pm$0.01 & 1.47$\pm$0.03 & 1.24$\pm$0.01\\
 24 & 12:19:25.284 & +05:47:54.66 & 1.39 & 2.38 & 0.97 & 85.1 & 24.84$\pm$0.03 & 24.17$\pm$0.06 & 23.14$\pm$0.01 & 0.77$\pm$0.02 & 1.01$\pm$0.02\\
... & ... & ... & ... & ... & ... & ... & ... & ... & ... & ... & ...\\
... & ... & ... & ... & ... & ... & ... & ... & ... & ... & ... & ...\\ 
   \hline
   \end{tabular}
  \end{minipage}
  \smallskip
 
  \begin{flushleft}
 
   $^{(1)}$ Globular Cluster ID \\
   $^{(2)}$ Equatorial Right Ascension (J2000) \\
   $^{(3)}$ Equatorial Declination (J2000) \\
   $^{(4)}$ Elongation (major to minor axis ratio)\\
   $^{(5)}$ FWHM assuming a Gaussian core, in units of WFPC2 pixels\\
   $^{(6)}$ SE{\tiny XTRACTOR} stellarity index (1 = point like; 0 = extended)\\
   $^{(7)}$ $\langle$S/N$\rangle$ ratio within aperture\\
   $^{(8)}$ Aperture-corrected $B$ band magnitude (if GC detected in $B$ band)\\
   $^{(9)}$ Aperture-corrected $V$ band magnitude\\
   $^{(10)}$ Aperture-corrected $I$ band magnitude\\
   $^{(11)}$ $B$-$V$ color (if GC detected in $B$ band)\\
   $^{(12)}$ $V$-$I$ color\\

  \end{flushleft}
 
 \end{table}
\end{landscape}


\appendix
\onecolumn
\begin{center}
 \section{INCOMPLETENESS SIMULATION}
 \label{INCOMPLETENESS SIMULATION}
\end{center}

In order to evaluate the incompleteness of the GCs samples 
(i.e., the fraction of GCs not detected due to faintness or issues related to the detection
process),
we set up an artificial source test to calibrate the results from SE{\footnotesize XTRACTOR}.
Simulated GCs were added to the NGC~4261 \emph{HST} mosaics and their
characteristics were measured with SE{\footnotesize XTRACTOR} using the same
setup as for the real data. The simulation was repeated several times in order
to improve the statistical results.
In the following, we describe the simulation scheme:

\bigskip

 \emph{Generating GC radial profiles.} We adopted the King model \citep[][equation 14]{kin62}
 to describe the GC brightness profile:

 $$I(r) = I_{c}({1\over{\sqrt{1+(r/r_{c})^2}}} -  {1\over{\sqrt{1+(r_{t}/r_{c})^2}}})^2$$

 where $I_{c}$ is the scale brightness, $r_{c}$ is the core radius
 (roughly similar to the half light radius $r_h$), $r_{t}$ is the tidal radius
 (which defines the radius up to which the gravitational field of the GC system can
 overcome the galactic gravitational field).
 The model can also be defined in terms of $r_{c}$ and the concentration parameter
 $c$ = $log_{10}(r_{t}/r_{c})$, which expresses the core extent in terms of the GC size.
 In our analysis, we adopted $r_{c}$ and $c$ as the parameters characterizing the model.
 Image templates reproducing King profiles, as would be seen
 at the distance of NGC~4261, were created spanning the typical range of
 $r_{c}$ ($1.5<r_{c}<10.0$~pc) and $c$ ($1.0<c<2.0$).
 We sampled from this 2-dimensional library of models when creating an artificial object.
 The $c$-space was sampled uniformly since catalogues of structural parameters for the
 GC populations in nearby galaxies show that the probability density distribution of $c$
 is almost flat between the imposed limits (e.g. \citealt{har96} for the Milky Way;
 \citealt{bar02} for M~31).
 We sampled the $r_{c}$-space following the $r_h$ distribution function
 measured by Jord{\'a}n for the GC population of early type galaxies in the Virgo cluster
 \citep[][equations 23 and 24]{jor05}.

 \medskip
 \emph{Generating instrument PSF.} Artificial PSFs for 
 different locations on each CCD of the \emph{HST} cameras were produced using the
 TinyTim tool \citep{kri95}, assuming a black body spectral
 distribution peaked at 3500~K (close to the typical colour temperatures of stars composing a GC). 
 In order to investigate the degree of variation of the PSFs across
 the WFPC2 field, we generated PSFs at the corners and the center of each CCD of WFPC2
 and we measured their Encircled Energy as a function of radius.
 The aperture used for the photometry on the real data encompasses 95\% of the PSF flux,
 with variations of only a few percent between different CCD locations or even different CCDs.
 Since these differences would not significantly affect the results of our simulation,
 we decided to adopt one single PSF model for the whole field, corresponding to the
 PSF at the central pixel of a WFC CCD.

 \medskip
 \emph{Defining source brightness.} A set of sources of different brightness was
 generated sampling between 26.0 and 19.0~mag (these limits were chosen to cover
 the whole range of observed magnitudes in each band).
 The sampling was slightly finer for fainter magnitudes,
 where most deviation between input and output parameters is expected.

 \medskip
 \emph{Adding artificial sources on the NGC~4261 mosaics.} For each iteration, a
 catalogue of uniformly distributed artificial source positions was generated.
 One random magnitude and one random King model
 from the set was assigned to each object position, without assuming any
 correlation between the size and the brightness of an artificial source.
 We convolved the King models with the instrument
 PSF and added the result on the NGC~4261 mosaic using the
 {\footnotesize IRAF} task \emph{mkobjects} 
 ({\footnotesize IRAF} package 2.13-BETA2, 2006).
 The spatial density of artificial objects was limited to 1 object every 10
 times the physical size of the simulated King model
 in order to avoid source confusion between the simulated objects themselves
 (each simulated object must be considered as an independent trial - many
 objects were simulated at the same time just to speed up the process).
 Simulated objects, though, could be cast onto real sources in the field,
 therefore accounting for real source confusion.
 The objects were uniformly distributed within the elliptical annulus constrained
 between the galactocentric radii $R_{GC}$ = 25$\arcsec$ (GC detection limit imposed
 by the background) and $R_{GC}$ = $R_{25}$ ($\sim$2$\arcmin$; galaxy major semi-diameter),
 as defined along the major axis.
 The axis ratio and P.A. parameters of the annulus have
 been obtained from a 2$D$ model of NGC~4261 (see $\S$\ref{GC Specific Frequency}).

 \medskip
 \emph{Artificial source photometry.} Each simulated field was analysed by
 SE{\footnotesize XTRACTOR} in order to obtain the aperture flux, position,
 axial ratio, FWHM and stellarity index of the artificial objects.
 The main aim of the simulation was to evaluate these parameters against
 the corresponding inputs as a function of their brightness.
 As a byproduct, we also got the incompleteness factors
 (i.e., the percentage of undetected sources) as a function of magnitude.

 \medskip
 \emph{Applying selection.} The criteria used to define the ``GC candidates''
 sample ($\S$\ref{Globular Cluster Selection}) were applied to the
 SE{\footnotesize XTRACTOR} catalogues, in order to reproduce the same selection
 as for the real data.

 \medskip
 \emph{Statistics.} For each iteration, we estimated the median of the
 measured values for each parameter, for the objects of a given input magnitude
 (the mean is not reliable since it is strongly driven by the outliers).
 At the end, the average of the medians of all the trials was selected as the expected
 value for the parameter. We estimated the dispersion around the expected value
 as the standard deviation of the medians.

\bigskip

The results of the simulation are discussed in $\S$\ref{Evaluating Incompleteness}
and plotted in Figure \ref{incompleteness} and \ref{FWHM}.

\newpage
\twocolumn

\label{lastpage}

\end{document}